\documentclass[twocolumn,aps,prb,showpacs,tightenlines,amsmath,amssymb]{revtex4}
\usepackage{graphicx}
\usepackage{amsmath}
\usepackage{amssymb}
\usepackage{mathrsfs}
\usepackage{bm}
\usepackage{colordvi}
\usepackage{epsfig}
\begin{document}              

\title{Kinetic theory of surface plasmon polariton in semiconductor nanowires}
\author{Y.\ Yin}
\email{yin80@ustc.edu.cn.}
\author{M.\ W.\ Wu}
\email{mwwu@ustc.edu.cn.}
\affiliation{Hefei National Laboratory for Physical Sciences at
  Microscale and Department of Physics, University of Science and
  Technology of China, Hefei, Anhui, 230026, China}

\date{\today}

\begin{abstract} 
  Based on the semiclassical model Hamiltonian of the surface plasmon polariton
  and the nonequilibrium Green-function approach, we present a microscopic
  kinetic theory to study the influence of the electron scattering on the
  dynamics of the surface plasmon polariton in semiconductor nanowires. The
  damping of the surface plasmon polariton originates from the resonant
  absorption by the electrons (Landau damping), and the corresponding damping
  exhibits size-dependent oscillations and distinct temperature dependence
  without any scattering. The scattering influences the damping by introducing a
  broadening and a shifting to the resonance. To demonstrate this, we
  investigate the damping of the surface plasmon polariton in InAs nanowires in
  the presence of the electron-impurity, electron-phonon and electron-electron
  Coulomb scatterings. The main effect of the electron-impurity and
  electron-phonon scatterings is to introduce a broadening, whereas the
  electron-electron Coulomb scattering can not only cause a broadening, but also
  introduce a shifting to the resonance. For InAs nanowires under investigation,
  the broadening due to the electron-phonon scattering dominates. As a result,
  the scattering has a pronounced influence on the damping of the surface
  plasmon polariton: The size-dependent oscillations are smeared out and the
  temperature dependence is also suppressed in the presence of the
  scattering. These results demonstrate the important role of the scattering on
  the surface plasmon polariton damping in semiconductor nanowires.
\end{abstract}


\maketitle

\section{INTRODUCTION}
\label{sec1}

Since the pioneering theoretical work by Ritchie\cite{Ritchie1957} and the
electron-loss spectroscopy measurements by Powell and Swan,\cite{Powell1960}
physics of surface plasmon polariton (SPP) has been extensively studied for more
than five decades.\cite{Chen1981, Wokaun1982, Rothen1988, Krebig1995,
  Zayats2005a, Pitarke2007, Vidal2010} SPPs are electromagnetic (EM) waves
coupled to the collective excitations of electrons on the surface of a
conductor.\cite{Raether1988, Barnes2003} In this coupling, the electrons
oscillate collectively in resonance with the EM waves and hence trap the EM
waves on the surface. The resonant coupling leads to the SPPs and gives rise to
their unique properties, such as the enhancement of the surface electric fields
and the slowing down of the group velocity of the EM waves.\cite{Ozbay2006,
  Kneipp1997, Nie1997, Xu1999, Hecht1996, Pendry2004, Sandtke2007, Maier2007,
  Zhang2008} Applications exploiting these properties have been widely studied
in biosensing,\cite{Homola1999} solar cells,\cite{Atwater2010} quantum
information processing,\cite{Chang2006, Akimov2007, Kolesov2009} subwavelength
optical imaging and waveguiding devices.\cite{Barnes2003, Zayats2005,
  Smolyaninov2005, Takahara1997, Cao2005, Archambault2010, Bergman2003}

The SPPs often suffer from dissipative losses.\cite{Bolt2011} Overcoming the
losses is crucial for the improvement of performance of many SPP-based devices,
such as the fidelity of the waveguide and the sensitivity of the single-molecule
sensor.\cite{Homola1999, Schuller2010} An effective modulation of the losses is
also highly desirable for active plasmonic devices proposed in recent
years.\cite{Krasavin2004, Nikolajsen2004, MacDonald2009} Thus, a thorough
understanding of the damping processes responsible for the dissipative losses is
essential. Since the SPPs in metals have been the major focus for many decades,
previous studies on the damping processes have traditionally been focused on
metals. It is found that the dominant damping process is the decay of the SPPs
into electrons, i.e., the Landau damping.\cite{Feibelman1982, Ekardt1985,
  Liebsch1987, Tsuei1989, Sprunger1992} The interband transitions and the
many-body exchange-correlations can have pronounced influences on the Landau
damping. These influences can be incorporated into microscopic models based on
time-dependent density-functional theory or semiclassical model Hamiltonian of
the SPP.\cite{Zaremba1987, Molina2002, Weick2005} Calculations based on these
models have shown good agreement with experiments.\cite{Silkin2004, Silkin2006,
  Gao2011}

In recent years, doped semiconductors, such as SiC, GaAs, InAs, Cu$_2$S and
Cu$_2$Se, are suggested as promising candidates to replace metals in SPP
applications.\cite{Hoffman2007, Taubner2006, Luther2011, Seletskiy2011,
  Vitiello2011} The SPPs in doped semiconductors are characterized by their
substantial low losses and tunable frequencies.\cite{Rivas2006, Rivas2006_1,
  Isaac2008, Naik2010} They are also easier to be manipulated via doping or
external electric/magnetic fields and to be integrated into complex, functional
circuits.\cite{Galli2006, Tang2008, Walters2010} The further development of the
SPPs in doped semiconductors requires a better understanding of their damping
processes. However, the physics involved in doped semiconductors can be quite
different from that in metals. For doped semiconductors, although the Landau
damping is still believed to be the leading damping process at large
wavevectors,\cite{Hasselbeck2008,comment1} the charge depletion/accumulation
layer\cite{Ehlers1986, Bell1996} and the electron scattering\cite{Bell2006,
  Abajo2010} are found to have important influence on the damping. Of particular
importance is the effect of the electron scattering, since the typical
scattering rate can be comparable to the SPP frequency in
semiconductors. However, to the best of our knowledge, this effect has only been
discussed by using phenomenological relaxation times.\cite{Bell2006, Abajo2010}
A microscopic theory exploiting this effect has not been established yet.

In this paper, by combining the semiclassical model Hamiltonian of the
SPP\cite{Zaremba1987, Molina2002, Weick2005, Gao2011} and the nonequilibrium
Green-function approach,\cite{Haug1996, mwu2010} we present a microscopic
kinetic theory to study the damping of the SPP in doped semiconductors, within
which the relevant electron scatterings are treated fully microscopically. The
main purpose of this work is to understand the influence of these scatterings on
the Landau damping of the SPP. To demonstrate this, we focus here on the SPPs in
InAs nanowires\cite{Chang2006, Seletskiy2011, Takahara1997, Cao2005} and
concentrate on the electron-impurity (ei), electron-phonon (ep) and
electron-electron (ee) Coulomb scatterings. We find that the scattering can have
pronounced influences on the Landau damping of the SPP by modulating the
resonance between the electrons and the SPPs. Different scattering has different
effect on the resonance. The main effect of the ei and ep scatterings is to
introduce a broadening to the resonance, whereas the ee scattering can not only
case a broadening, but also introduce a shifting to the resonance. For InAs
nanowires, the ep-scattering--induced broadening is found to be the dominant
effect. These effects can lead to a pronounced influence on the damping of the
SPP, which can be seen from both the size and temperature dependence of the SPP
damping: (1) The size-dependent oscillations of the SPP damping are smeared out,
and (2) the temperature dependence of the SPP damping is suppressed by the
scattering. These results demonstrate the important role of the electron
scattering on the SPP damping.

This paper is organized as follows. In Sec.~\ref{sec2}, we introduce the
semiclassical model Hamiltonian of the SPP for semiconductor nanowires and
briefly outline the derivation of the kinetic equations. We also present an
analytic solution for the SPP damping which provides a simple and physically
transparent picture to understand the influence of the scattering on the SPP
damping process. In Sec.~\ref{sec5}, we discuss in detail the influence of the
electron scattering on the SPP damping by numerical solving the kinetic
equations. The importance of the scattering is demonstrated by studying its
influences on the temperature and size dependence of the SPP damping. The
analytic solution is also compared with the numerical ones in this section. We
summarize and discuss in Sec.~\ref{sec8}.

\section{Model and Formalism}
\label{sec2}

\subsection{Semiclassical model for SPP-electron system}
\label{sec2_0}

\begin{figure}
  \centering
  \includegraphics[width=6.5cm]{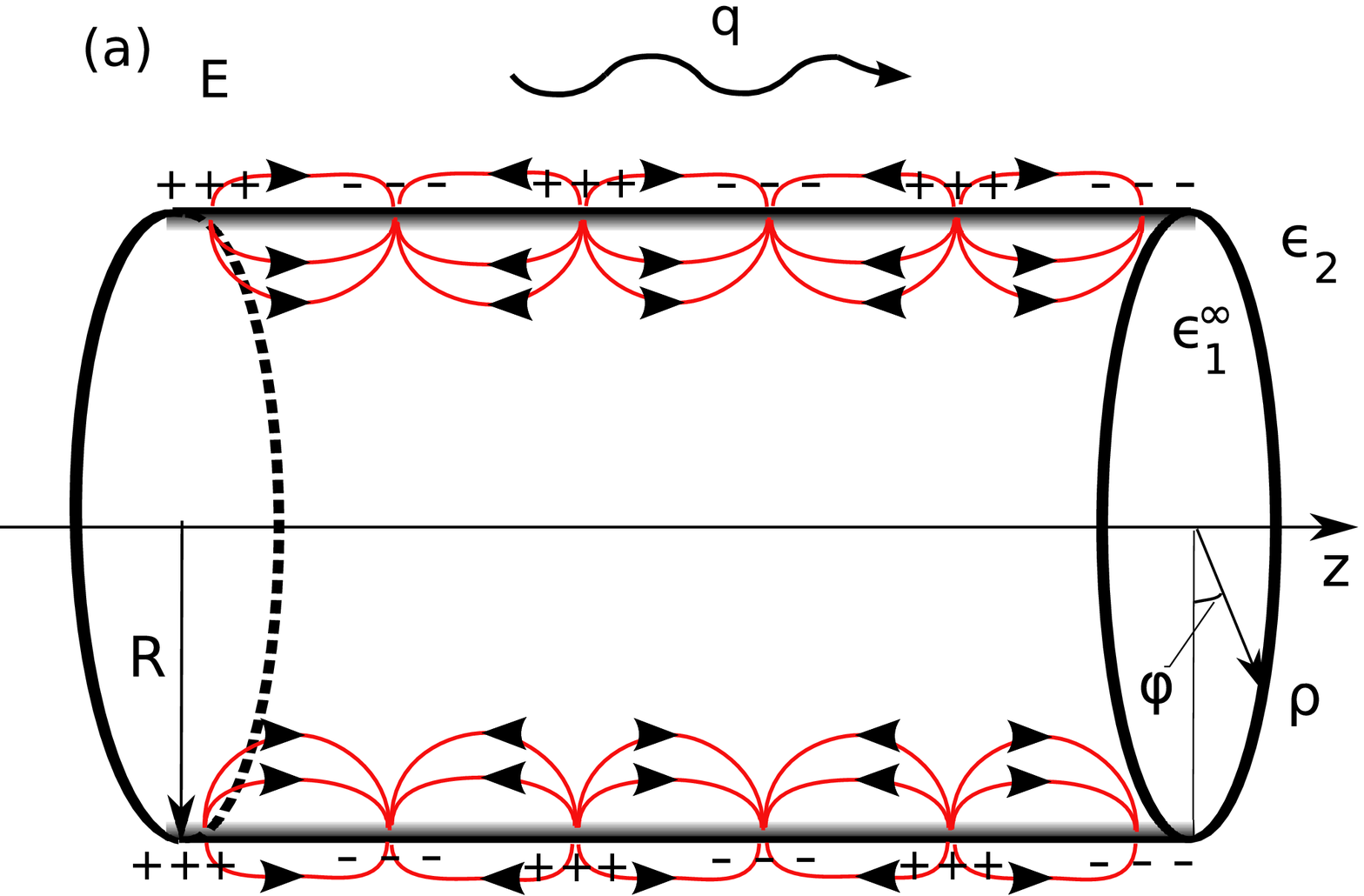}\\
  \includegraphics[width=6.5cm]{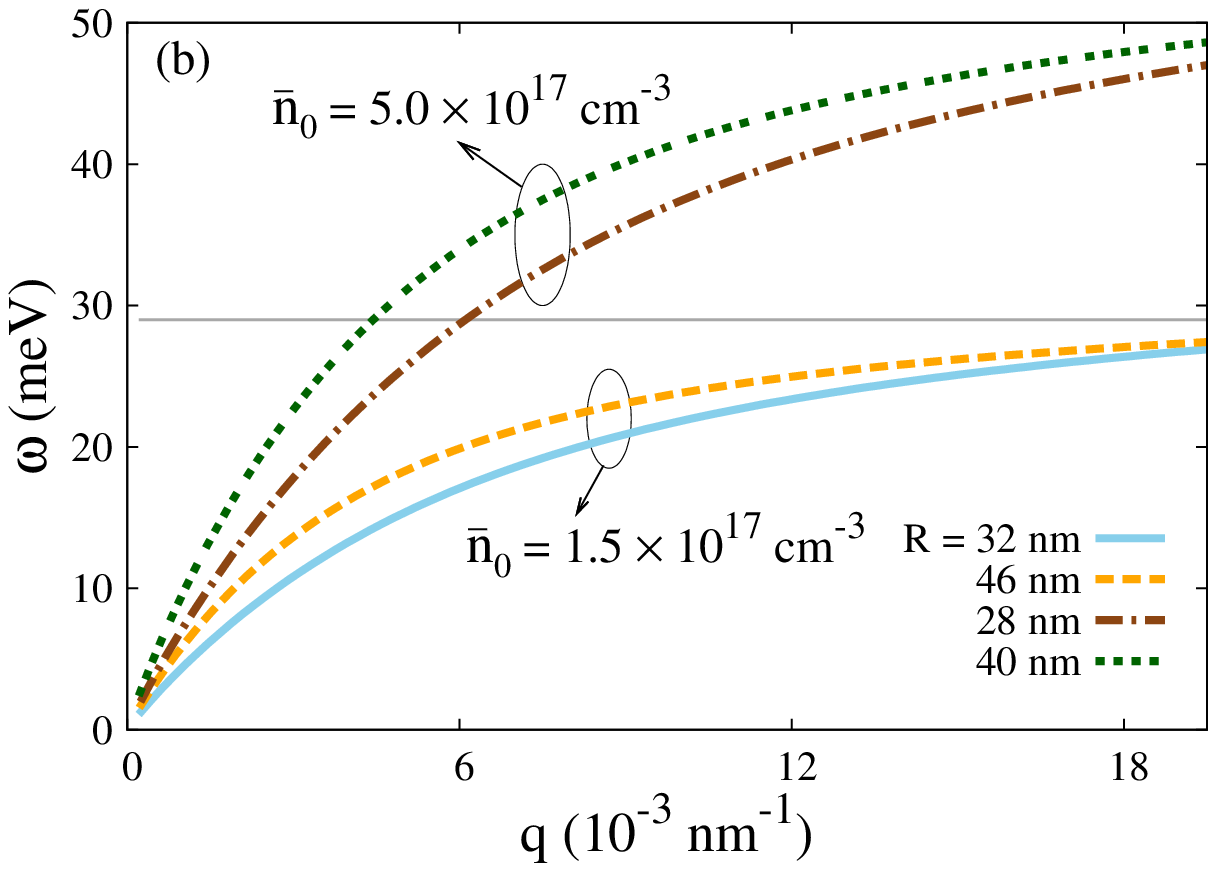}
  \caption{(Color online) (a) Schematic of the structure of the electron field
    $E$ and corresponding surface charge of the axial symmetric SPP mode in a
    cylindrical nanowire. $\epsilon^{\infty}_1$ ($\epsilon_2$) is the dielectric
    constant inside (outside) the nanowire. The red curves with arrows represent
    the electric fields of the SPP mode. (b) The dispersive relation of the
    axial symmetric SPP mode for nanowires with different electron density
    $\bar{n}_0$ and wire radius $R$. The thin grey line marks the energy of the
    LO phonon.}
  \label{cl:fig1}
\end{figure}

We consider an $n$-type free-standing cylindrical nanowire with radius $R$ as
illustrated in Fig.~\ref{cl:fig1}(a). The $z$-axis is chosen to be along the
wire. Following the semiclassical approach developed in previous
works,\cite{Zaremba1987, Molina2002, Weick2005, Gao2011} we decompose the
Hamiltonian into
\begin{equation}
  H = H_{\rm SPP} + H_{\rm el} + H_{\rm SPP\text{-}el}, \label{cl:eq0}
\end{equation}
where $H_{\rm SPP}$, $H_{\rm el}$ and $H_{\rm SPP\text{-}el}$ are Hamiltonians
for the SPP, electrons and the SPP-electron coupling, respectively. Here we only
present the Hamiltonian, leaving the details to Appendix~\ref{app1}.

For nanowires, there exists one fundamental SPP mode with axial symmetry, which
has no cutoff at low frequency and has been found to be important for both the
terahertz emission and quantum subwavelength optics.\cite{Chang2006,
  Seletskiy2011, Takahara1997, Cao2005} In this paper, we focus on the dynamics
of this SPP mode. The corresponding Hamiltonian takes the form
\begin{eqnarray}
  \label{cl:eq1}
  H_{\rm SPP} & = & \sum_q \Omega_q b^{\dagger}_q b_q,
\end{eqnarray}
where $b_q$($b^{\dagger}_q$) represents the annihilation(creation)
boson operator for the SPP, with $\Omega_q$ being the corresponding dispersive
relation illustrated in Fig.~\ref{cl:fig1}(b). Note that we set $\hbar=1$
throughout this paper.

The Hamiltonian of electrons can be written as
\begin{equation}
  \label{cl:eq2}
  H_{\rm el} = H^0_{\rm el} + H_{\rm ei} + H_{\rm ep} + H_{\rm ee},  
\end{equation}
where the free-electron Hamiltonian $H^0_{\rm el}$ is modelled by a mean-field
potential. The interaction Hamiltonians $H_{\rm ei}$, $H_{\rm ep}$ and $H_{\rm
  ee}$ represent the ei, ep and ee interaction, respectively.

By choosing the mean-field potential to be an infinite cylindrical potential
well with radius $R$, the free-electron Hamiltonian $H^{0}_{\rm el}$ can be
written as
\begin{eqnarray}
  \label{cl:eq3}
  H^{0}_{\rm el} & = & \sum_{n k \sigma} \varepsilon^n_k c^{\dagger}_{n k \sigma} c_{n k \sigma},
\end{eqnarray}
in which $\varepsilon^n_k = \frac{k^2 + (\lambda^{\tilde{m}}_{\tilde{n}}/R)^2}{2
  m^{\ast}}$ is the eigenenergy with $m^{\ast}$ representing the electron
effective mass. The composed index $n=(\tilde{m}, \tilde{n})$ labels the
electron subband with $\tilde{m}$ and $\tilde{n}$ representing the angular and
radial quantum numbers, respectively. $\lambda^{\tilde{m}}_{\tilde{n}}$ denotes
the $\tilde{n}$-th zero of the Bessel function of the first kind
$J_{\tilde{m}}(x)$. The corresponding eigenstates read
\begin{eqnarray}
  \label{cl:eq6}
  \psi_{n k}(\rho, \varphi, z) & = & \frac{J_{\tilde{m}}(\lambda^{\tilde{m}}_{\tilde{n}} \rho/R)}{\sqrt{\pi R} J_{\tilde{m}+1}(\lambda^{\tilde{m}}_{\tilde{n}})} e^{i \tilde{m} \varphi}
  e^{i k z}.
\end{eqnarray} 
The ei interaction Hamiltonian can be written as
\begin{eqnarray}
  H_{\rm ei} & = & \sum^{N_i}_i \sum_{k q \sigma} v^{n n'}_q \rho_i(q)
  c^{\dagger}_{n' k+q \sigma} c_{n k \sigma}, \label{cl:eq4}  
\end{eqnarray}
with $N_i$ being the total impurity number and $\rho_i(q) = e^{-i q z_i}$. Here
we have assumed that the impurities are distributed on the surface of the
nanowire with an axially symmetric distribution. $v^{nn'}_q$ is the matrix
element for the ei interaction. The ep interaction Hamiltonian can
be written as
\begin{eqnarray}
  H_{\rm ep} & = & \sum_{\bm{Q} q} \sum_{n n' k \sigma} M^{nn'}_{\bm{Q} q} \Big( a_{\bm{Q} q} +
  a^{\dagger}_{-\bm{Q} -q} \Big) c^{\dagger}_{n k \sigma} c_{n' k-q \sigma},   \label{cl:eq5}
\end{eqnarray}
where $a_{\bm{Q} q}$($a^\dagger_{\bm{Q} q}$) represents the
annihilation(creation) operator for the LO phonons, with $\bm{Q}$ and $q$
representing the components of the phonon momentum perpendicular and parallel to
the nanowire. Here we use bulk phonons in the present investigation, which is
valid for nanowires with large diameters.\cite{Konar2007, Hormann2011}
$M^{nn'}_{\bm{Q} q}$ is the matrix element for the ep interaction. It is noticed
that although surface-optical (SO) phonons\cite{Hormann2011} also exist in
nanowires, they are of marginal importance since the corresponding
electron-SO-phonon interaction is rather weak compared with the
electron-LO-phonon interaction for the nanowires we consider here. The influence
of the SO phonons will be further addressed in Sec.~\ref{sec5_2}.  The ee
interaction Hamiltonian can be written as
\begin{eqnarray}
  H_{\rm ee} & = & \sum_{k k' q} \sum_{n n'} \sum_{\sigma \sigma'}V_{q}^{nn'} c_{n k \sigma}^{\dagger}c_{n' k' \sigma'}^{\dagger}c_{n'
    k'+q \sigma'}c_{n k-q \sigma}, \label{cl:eq8}
\end{eqnarray}
where $V^{nn'}_q$ is the matrix element for the ee interaction. The
SPP-electron coupling Hamiltonian $H_{\rm SPP\text{-}el}$ can be written as
\begin{equation}
  \label{cl:eq7}
  H_{\rm SPP\text{-}el} = \sum_{n k \sigma} \sum_{n' k'} g^{nn'}_{k-k'}
  \Big( b_{k-k'} + b^{\dagger}_{k'-k} \Big) c^{\dagger}_{n k \sigma}  c_{n' k' \sigma},
\end{equation}
where $g^{nn'}_q$ is the SPP-electron coupling matrix element. In these
equations, matrix elements $v^{nn'}_q$, $M^{nn'}_{\bm{Q} q}$,
$V^{nn'}_q$ and $g^{nn'}_q$ are given in detail in Appendix~\ref{app1}.

\subsection{Kinetic equations}
\label{sec2_1}

In this section, we briefly outline the derivation of the kinetic equations for
the SPP-electron system. The details can be found in Appendix~\ref{app2}.

The damping of the SPP is obtained by studying the temporal evolution of a
coherent SPP wave packet with central wavevector $Q_s$, which can be expressed
as $| B_s \rangle = \sum_q p^{Q_s}_q e^{-\frac{1}{2} |B_q|^2} e^{b_q
  b^{\dagger}_q} | 0 \rangle$ (Ref.~\onlinecite{Roudon2000}). The line-shape
function of the wave packet $p^{Q_s}_q$ is chosen to be $p^{Q_s}_q = \frac{
  \sin[ ( Q_s - q ) L/2 ] }{ ( Q_s - q ) L/2 }$, where $L$ is the wave packet
length. Such wave packet is typical in a Fabry-Perot SPP resonator, which has
been observed in various SPP systems.\cite{Kolesov2009, Ditlbacher2005,
  Miyazaki2006, Cao2010}

The amplitude of the wave packet can be described by $B_s = \sum_q p^{Q_s}_q
B_q$. The kinetic equation of $B_s$ is obtained from the Heisenberg equation of
the SPP annihilation operator $b_q$, which has the form
\begin{eqnarray}
  \partial_t B_s(t) & = & \sum_{nn', kk', \sigma} p^{Q_s}_{k-k'}  g^{nn'}_{k-k'} G^{<}_{\sigma}(n'k', nk; tt),
  \label{eom:eq1}
\end{eqnarray}
where $G^{<}_{\sigma}(nk, n'k'; tt')$ is the ``lesser'' electron Green
function defined as $G^{<}_{\sigma}(nk, n'k'; tt') = i \langle
c^{\dagger}_{n' k' \sigma}(t') c_{n k \sigma}(t) \rangle$
(Ref.~\onlinecite{Haug1996}).

The kinetic equation of the electron Green function $G^{<}_{\sigma}(nk, n'k'; tt)$ is derived by using the nonequilibrium Green-function
approach, which can be written as
\begin{eqnarray}
  &&\hspace{-1cm} \Big[ -i \partial_t - ( \varepsilon^{n'}_{k'} - \varepsilon^n_k ) \Big]
  G^{<}_{\sigma}(nk, n'k'; tt) \nonumber\\
  &&\hspace{0.0cm} = \sum_{\bar{n} q} ( B_{-q} + B^{\dagger}_q ) \Big[
  g^{\bar{n} n'}_q G^{<}_{\sigma}(nk, \bar{n}k'-q; tt) \nonumber\\
  &&\hspace{0.0cm}\mbox{} - g^{n \bar{n}}_q G^{<}_{\sigma}(\bar{n}k+q, n'k'; tt)
  \Big] + I^{< \sigma}_{nk, n'k'}(t),
  \label{eom:eq3}
\end{eqnarray}
where the first term in the right hand side of the equations is the coherent
driving term of the SPP, while the second term is the scattering term consisting
the ei, ep and ee scatterings.

Within the rotating wave approximation relative to the SPP central frequency
$\Omega_s$,\cite{Haug1996, mwu2010} we obtain the kinetic equations for the
SPP-electron system
\begin{eqnarray}
  &&\hspace{-0.75cm}\partial_t \bar{B}_s(t) = -i \sum_{nn' \sigma} \sum_{k k'} p^{Q_s}_{k-k'}  g^{nn'}_{k-k'} \Big[ P_{\sigma}(nk, n'k'; t) \Big]^{\dagger}, \label{eom:eq4_1}\\
  &&\hspace{-0.75cm}\partial_t P_{\sigma}(nk, n'k'; t) = i \omega^{nn'}_{kk'} P_{\sigma}(nk, n'k'; t) \nonumber\\
  &&\hspace{-0.25cm}{}+ i g^{nn'}_{k-k'} p^{Q_s}_{k-k'} \bar{B}^{\dagger}_s
  \Big[ f_{n \sigma}(k) - f_{n' \sigma}(k') \Big] + \bar{I}^{\sigma}_{nk, n'k'}, \label{eom:eq4_2}
\end{eqnarray}
with the detuning
\begin{equation}
  \label{eom:eq4_3}
  \omega^{nn'}_{kk'}=\varepsilon^{n'}_{k'} - \varepsilon^n_k - \Omega_s.
\end{equation}
In the above equations, $\bar{B}_s(t) = B_s(t) e^{i \Omega_s t}$ and $P_{
  \sigma}(nk, n'k'; t) = -i G^{<}_{ \sigma}(nk, n'k'; tt) e^{i \Omega_s t}$
represent the SPP amplitude and electron polarization, respectively. $f_{
  n\sigma}(k)$ represents the equilibrium electron distribution which is
conventionally chosen to be the spin-unpolarized Fermi-Dirac distribution at
temperature $T$.

It should be emphasized that in the derivation, we have treated the SPP-electron
coupling $g^{ nn'}_q$ perturbatively and linearized the equations by keeping
only terms up to the linear order of $g^{ nn'}_q$. Thus the SPP couples only
to the electron polarization corresponding to the off-diagonal electron Green
function with respect to the electron momentum $k$ and subband index $n$.

The scattering term within the rotating wave approximation can be expressed as
$\bar{I} = \bar{I}^{\rm ei} + \bar{I}^{\rm ep} + \bar{I}^{\rm ee}$, where
$\bar{I}^{\rm ei}$, $\bar{I}^{\rm ep}$ and $\bar{I}^{\rm ee}$ are contributions
from the ei, ep and ee scatterings, respectively. Their expressions can be found
in Appendix~\ref{app2}.

\subsection{Landau damping process}
\label{sec2_2}

In the kinetic equations, Eq.~\eqref{eom:eq4_2} describes the resonant
excitation of the electron polarizations, while Eq.~\eqref{eom:eq4_1} describes
the back-action of the polarizations to the SPP. The summation in
Eq.~\eqref{eom:eq4_1} implies that even without any scattering, the phase-mixing
between polarizations with different frequencies can also lead to the damping of
the SPP, which is the origin of the Landau damping.

To further clarify the Landau damping process described by the kinetic
equations, we solve the equations without the scattering
term $\bar{I}$. From Eq.~\eqref{eom:eq4_2}, the corresponding polarization can
be solved as
\begin{eqnarray}
  &&\hspace{-1.1cm}P_{\sigma}(nk, n'k'; t) = -\delta_{ k'-k-q} g^{ nn'}_{ q} p^{Q_s}_{ q}
  \bar{B}^{\dagger}_s(t) \nonumber\\
  &&\hspace{-1.1cm}{} \times [ f_{n\sigma}(k) - f_{n'\sigma}(k') ] [ e^{i (\omega^{ nn'}_{ kk'} + i 0^{+}) t}
  - 1 ]/(\omega^{ nn'}_{ kk'} + i 0^{+} ). \label{eom2:eq1}
\end{eqnarray}
Note that in the derivation, we have assumed that the SPP amplitude $\bar{B}_s$
varies slowly compared to the polarization $P$.

By substituting Eq.~\eqref{eom2:eq1} into Eq.~\eqref{eom:eq4_1} and taking the
long time limit $t \to \infty$, one gets
\begin{equation}
  \label{eom2:eq2}
  \partial_t \bar{B}_s/\bar{B}_s = - ( \tau^{-1} + i \omega_s ),
\end{equation}
where $\omega_s$ represents the frequency shifting, expressed as
\begin{equation}
  \omega_s = \sum_{ nk, n'k', q \sigma} | p^{ Q_s}_{ q}
  g^{ nn'}_{ q} |^2 [ f_{ n'\sigma}(k') - f_{ n\sigma}(k) ]
  \frac{1}{\omega^{ nn'}_{ kk'}} \delta_{ k'-k-q},
  \label{eom2:eq3}
\end{equation}
in which the summation is understood as a principal value integral. $\tau^{-1}$
is the damping rate of the SPP, which has the form
\begin{equation}
  \frac{1}{\tau} = \pi \sum_{ nk, n'k', q \sigma} | p^{
    Q_s}_{ q} g^{ nn'}_{ q} |^2 [ f_{ n'\sigma}(k') - f_{
    n\sigma}(k) ] \delta(\omega^{ nn'}_{ kk'}) \delta_{ k'-k-q}.
  \label{eom2:eq4}
\end{equation}
Note that Eq.~\eqref{eom2:eq4} agrees with the Landau damping rate derived from
the Fermi golden rule in the literature.\cite{Weick2005, Gao2011}

\begin{figure}
  \centering
  \includegraphics[width=5.5cm]{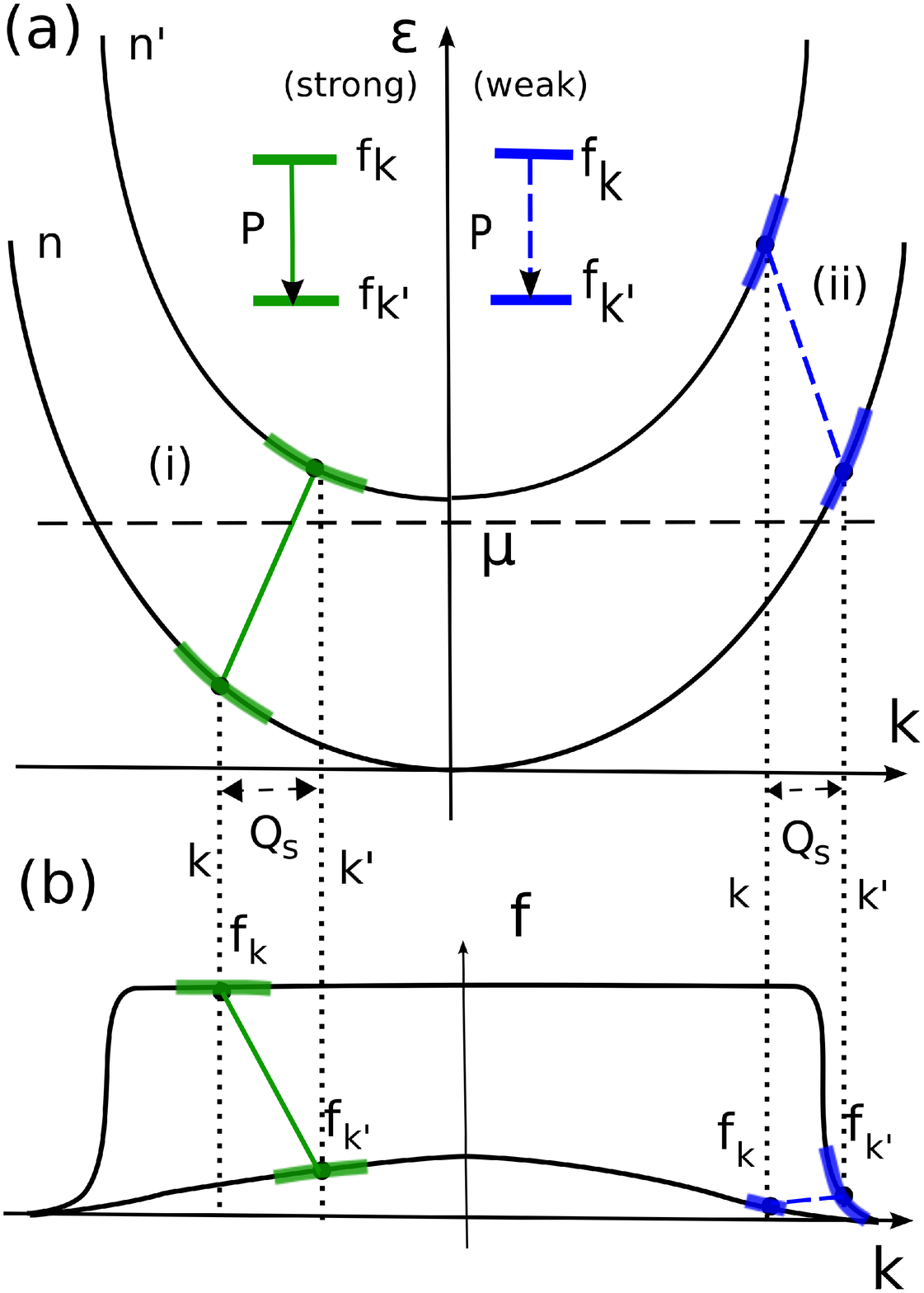}
  \includegraphics[width=5.5cm]{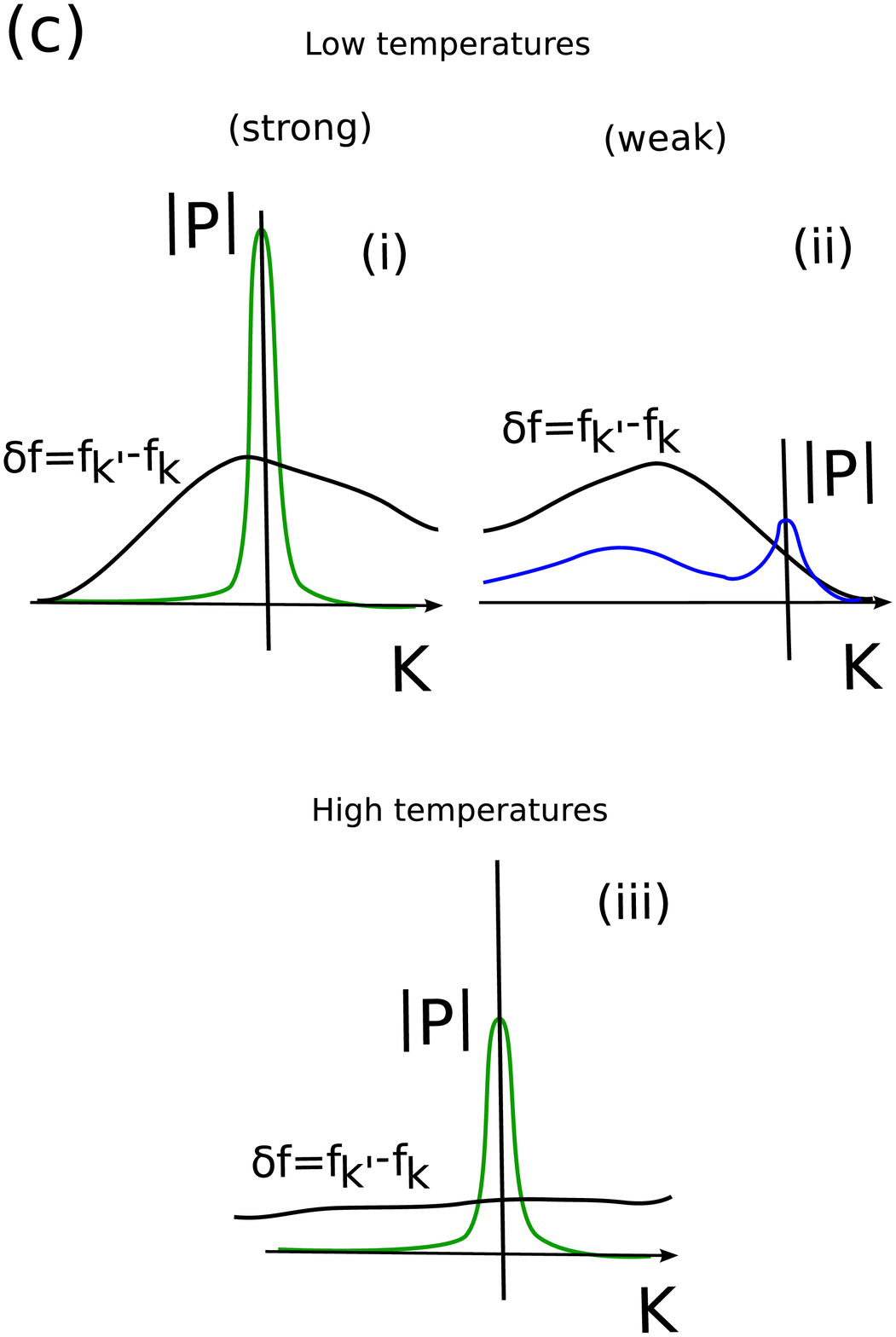}
  \caption{(Color online) Schematic of the resonant pairs corresponding to the
    case of (i) strong Landau damping and (ii) weak Landau damping in the
    electron spectrum (a) and in the electron distribution (b). The two
    wavevector regions of the resonant pairs are illustrated by thick green/blue
    curves corresponding to the strong/weak Landau damping regime. The resonant
    pairs are centralized around the resonance corresponding to the SPP central
    wavevector $Q_s$ as indicated by the vertical black dotted curves. The
horizontal dashed line marks the chemical potential of the electrons
$\mu$. (c) The electron polarization and electron 
population difference as function of
    the center wavevector $K=\frac{k+k'}{2}$ with $k'-k=Q_s$. (i)/(ii)
    corresponds to the strong/weak Landau damping regime at low temperatures and
    (iii) corresponds to both the strong and the weak Landau damping regimes at
    high temperatures. The vertical solid line marks the resonance corresponding
    to $Q_s$.}
  \label{eom:fig1}
\end{figure}

The above solution suggests that the Landau damping process can be understood as
the resonant absorption of the SPP by electrons. The two $\delta$-functions in
Eq.~\eqref{eom2:eq4} indicate that for a monochromatic SPP wave with wavevector
$q$, the absorption occurs between pairs of states $|nk\rangle$ and
$|n'k'\rangle$ satisfying the energy and momentum conservations
\begin{eqnarray}
  \omega^{ nn'}_{ kk'} & = & 0, \label{eom:eq_11}\\
  k' - k & = & q. \label{eom:eq_1}
\end{eqnarray}
Each pair of the states $|nk\rangle$ and $|n'k'\rangle$ consist a resonant pair
$(nk, n'k')$ relevant for the SPP damping. For the multi-subband system, there
usually exist several such resonant pairs, laying between different subbands $n$
and $n'$ and being well-separated from each other. For a nonmonochromatic SPP
wave packet with sufficiently narrow spectrum in $q$, the two states
$|nk\rangle$ and $|n'k'\rangle$ of each resonant pair become two wavevector
regions. Note that in the following discussion, we shall focus on the
nonmonochromatic SPP wave packet, and the resonant pair is referred to as the
wavevector region unless otherwise specified. The resonant pairs are illustrated
in Figs.~\ref{eom:fig1}(a) and (b). By using the resonant pairs, one can rewrite
Eq.~\eqref{eom2:eq4} into
\begin{eqnarray}
  \tau^{-1} & = & \sum_{ i} \tau^{-1}_{ i}, \label{eom2:eq5_1}\\
  \tau^{-1}_{ i} & = & \pi \sum_{ q \sigma} \sum_{ (nk, n'k') \in i} |p^{
    Q_s}_{ q} g^{ nn'}_{ q}|^2 [ f_{ n'\sigma}(k') - f_{
    n\sigma}(k) ] \nonumber\\
  && {} \times \delta(\omega^{ nn'}_{ kk'}) \delta_{
    k'-k-q}, \label{eom2:eq5_2}
\end{eqnarray}
with $i$ being the index for the resonant pair corresponding to the SPP wave
packet, whose spectrum is decided by the line-shape function $p^{
  Q_s}_q$. $(nk, n'k') \in i$ means that the two states $|nk\rangle$ and
$|n'k'\rangle$ belong to the $i$-th resonant pair. Thus one can see that the
damping rate of the SPP wave packet is the sum over the absorption rates of all
the relevant resonant pairs.
 
Note that the electron population difference $\delta f = f_{ n'\sigma}(k') - f_{
  n\sigma}(k)$ of the corresponding resonant pair plays an important role on its
contribution to the SPP damping. For the degenerate electrons where a
well-defined Fermi surface exists around the chemical potential, there exist two
regimes of the SPP damping: (i) a strong Landau damping regime where states
$|nk\rangle$ and $|nk'\rangle$ of a resonant pair lay in each side of the
chemical potential in the electron spectrum, leading to a large population
difference $\delta f$ and hence a strong SPP damping; (ii) a weak Landau damping
regime where the chemical potential lies outside all the resonant pairs, leading
to a small $\delta f$ and a weak SPP damping. This is illustrated in
Figs.~\ref{eom:fig1}(a) and (b). Note that at high enough temperatures, the
Fermi surface can be smeared out and such difference vanishes.

It should be emphasized that according to Eq.~\eqref{eom2:eq1}, the resonant
pair can be visualized as the resonant peak in the polarizations between the two
subbands $n$ and $n'$. In Fig.~\ref{eom:fig1}(c), we illustrate the
polarizations $P(nk, n'k')$ corresponding to the resonant pairs shown in
Figs.~\ref{eom:fig1}(a) and (b) as function of the center wavevector
$K=(k+k')/2$ with $k'-k=Q_s$. One finds that the polarization exhibits a
Lorentzian peak around the resonance corresponding to the central wavevector
$Q_s$, as indicated by Eq.~\eqref{eom2:eq1}.

Note that such resonant peak can show different features in the strong and weak
Landau damping regimes at low temperatures. In the strong Landau damping regime,
the corresponding polarization $P$ exhibits a strong resonant peak concentrated
in the region with large $\delta f$ [(i) in Fig.~\ref{eom:fig1}(c)], indicating
a large SPP absorption by the electrons. In contrast, in the weak Landau damping
regime, the corresponding resonant peak is weak and lies outside the large
$\delta f$ region [(ii) in Fig.~\ref{eom:fig1}(c)], indicating a small SPP
absorption. In addition to the resonant peak, side peaks can also exist in the
off-resonant regime due to the corresponding large $\delta f$. At high
temperatures where the Fermi surface is smeared out, the population difference
is rather flat for both the strong and weak Landau damping regimes and the
corresponding polarization exhibits a strong peak around the resonance for both
regimes [(iii) in Fig.~\ref{eom:fig1}(c)].

\subsection{Influence of the scattering}
\label{sec2_3}

The scattering influences the Landau damping by changing the resonant excitation
of the polarizations. Specifically, (1) the scattering can introduce dissipative
channels, inducing a decay of the polarization; (2) the scattering between
polarizations with different precession frequencies induces a frequency-mixing,
leading to a modification of the polarization precession frequency. These two
effects can be further clarified by assuming that for each resonant pair, the
scattering term has the form
\begin{equation}
  \bar{I}^{ i \sigma}_{ nk,n'k'} = \sum_q \Gamma_i [ P_{\sigma}(nk-q, n'k'-q) - P_{\sigma}(nk, n'k') ],
  \label{eom3:eq0}
\end{equation}
where $\Gamma_i$ stands for the phenomenological relaxation rate for the
polarization of the $i$-th resonant pair. Note that $(nk, n'k')$ and $(nk-q,
n'k'-q)$ belong to the $i$-th resonant pair.

For each resonant pair, the polarization $P$ in the
scattering term $\bar{I}$ given above can be obtained by treating $\Gamma_i$
perturbatively and solving Eq.~\eqref{eom:eq4_2} order by order, yielding
\begin{eqnarray}
  \label{eom3:eq1}
  P_{\sigma}(nk, n'k';t) & = & \delta_{ k'-k-q} g^{ nn'}_{ q} p^{Q_s}_{ q}
  \bar{B}^{\dagger}_s(t) [ f_{n\sigma}(k) - f_{n'\sigma}(k') ] \nonumber\\
  &&\hspace{-2.5cm}{} \times [ e^{i
    (\omega^{ nn'}_{ kk'} - \bar{\Gamma}^{a}_i + i \bar{\Gamma}^{b}_i)
    t} - 1] /(\omega^{ nn'}_{ kk'} - \bar{\Gamma}^{a}_i + i
  \bar{\Gamma}^{b}_i),  
\end{eqnarray}
where
\begin{eqnarray}
  &&\hspace{-1cm} \bar{\Gamma}^b_i = \Gamma_i \sum_q ( 1 -
  \frac{ \omega^{ nn'}_{ kk'} }{\omega^{ nn'}_{ k-q, k'-q} } ), \label{eom3:eq1_1}\\
  &&\hspace{-1cm} \bar{\Gamma}^a_i = \pi \Gamma_i \sum_q (\omega^{ nn'}_{
    kk'} - \omega^{ nn'}_{ k-q, k'-q}) \delta(\omega^{ nn'}_{ k-q, k'-q}). \label{eom3:eq1_2}
\end{eqnarray}
The summation in Eq.~\eqref{eom3:eq1_1} is understood as a principal value
integral. Note that we have omitted the $k,k'$-dependence of
$\bar{\Gamma}^{a(b)}_i$ inside each resonant pair for simplicity. The detail of
the derivation is given in Appendix~\ref{app3}.

\begin{figure}
  \centering
  \includegraphics[width=5.5cm]{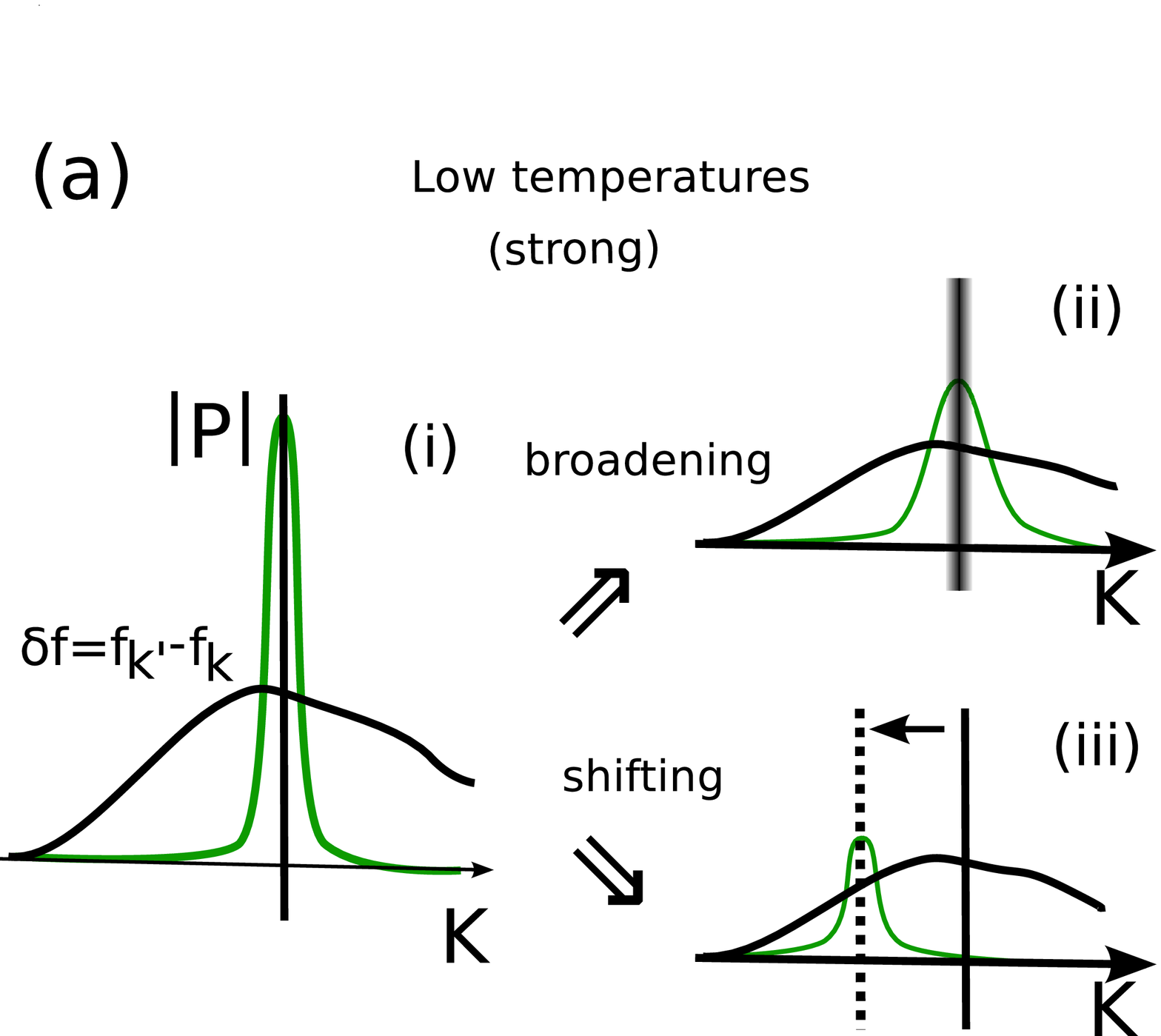}
  \includegraphics[width=5.5cm]{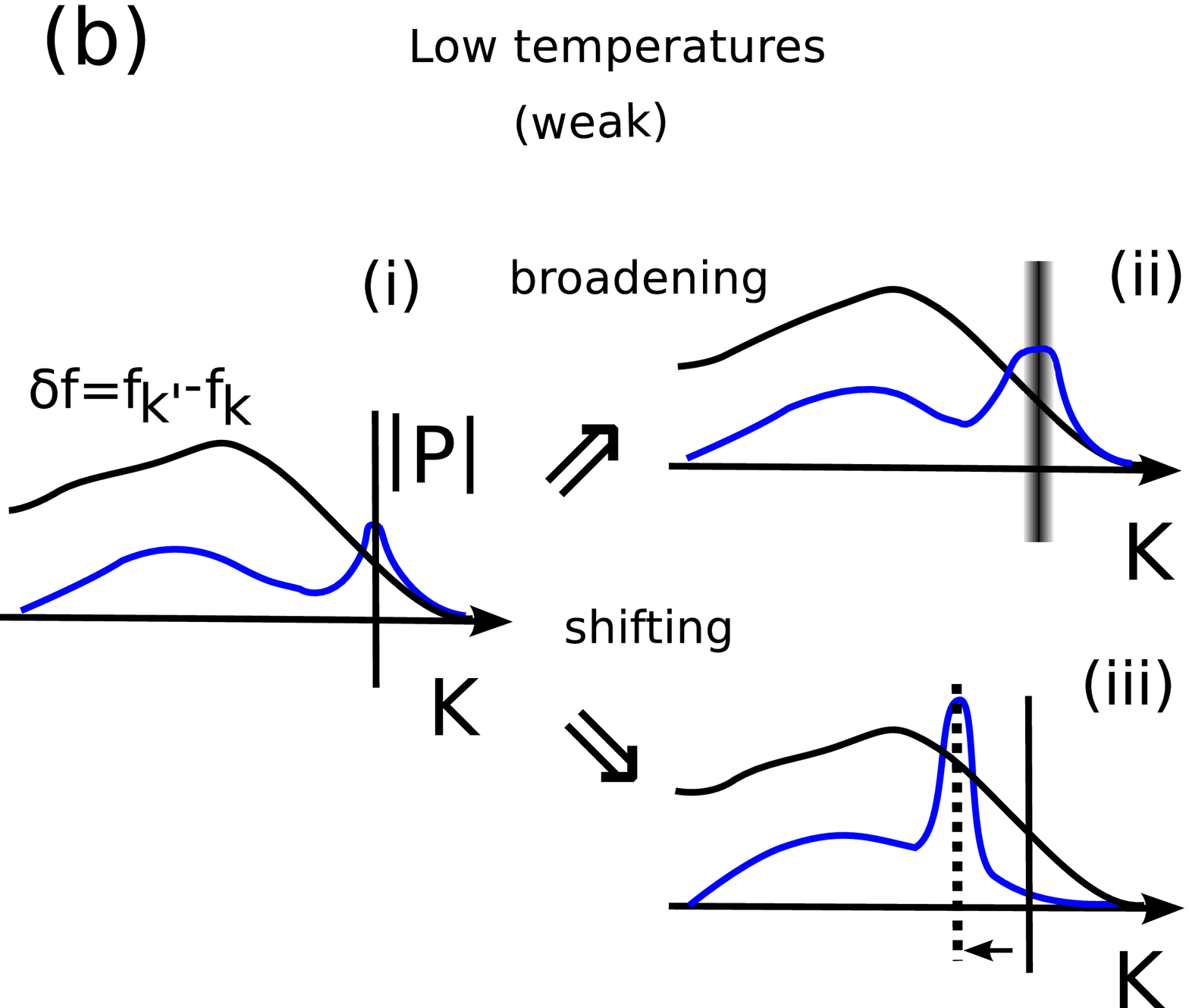}
  \includegraphics[width=5.5cm]{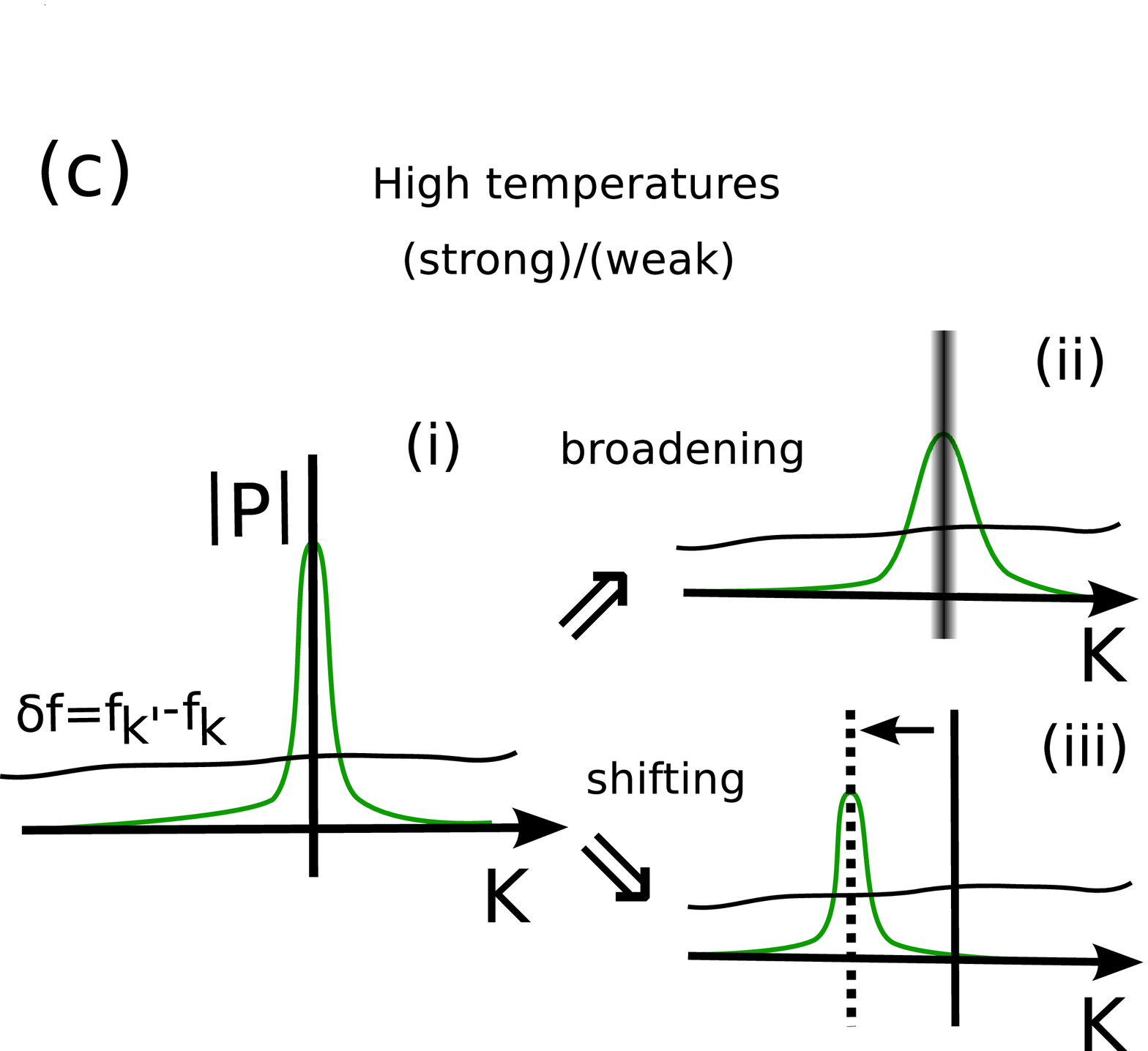}
  \caption{(Color online) Schematic of the effect of the broadening and
    shifting on the electron polarization for (a) strong Landau damping regime
    at low temperatures; (b) weak Landau damping regime at low temperatures and
    (c) both strong and weak Landau damping regimes at high temperatures. The
    thin solid vertical lines represent the resonance without broadening and
    shifting. The resonances with the broadening and shifting are represented
by the thick solid and thin dotted vertical lines,
    respectively. For clarification, only the shifting towards the small $K$
    direction is illustrated. Note that the population differences $\delta f$
    become flat in both regimes at high temperatures.}
  \label{eom:fig2}
\end{figure}

By comparing Eq.~\eqref{eom3:eq1} to Eq.~\eqref{eom2:eq1}, one can see that the
detuning $\omega^{ nn'}_{ kk'}$ in the resonant denominator is modified
into $\omega^{ nn'}_{ kk'} -\bar{\Gamma}^{a}_i$, indicating that the
polarization precession frequency is shifted by the scattering. The scattering
also induces a finite imaginary part $\bar{\Gamma}^{b}_i$ to the resonant
denominator, representing the decay of the polarization due to the scattering.

The above solution indicates that the scattering modifies the resonance between
the polarization and the SPP by introducing both an energy shift and an energy
broadening to the corresponding resonance pairs. On one hand, the
energy shift modifies the energy conservation Eq.~\eqref{eom:eq_11} into
\begin{equation}
  \label{eom3:eq4}
  \omega^{ nn'}_{ kk'} - \bar{\Gamma}^a_i = 0.
\end{equation}
Thus the corresponding resonant pairs are shifted by the scattering. On the
other hand, the energy broadening loosens the energy conservation constraint
given by Eq.~\eqref{eom:eq_11}. Thus the corresponding resonant pairs are
broadened. Note that the broadening and shifting are usually small and cannot
induce overlaps between different resonant pairs.

Accordingly, the broadening $\bar{\Gamma}^{b}_i$ and shifting
$\bar{\Gamma}^{a}_i$ also manifest themselves in the SPP damping rate. Following
the same procedure of the derivation of Eq.~\eqref{eom2:eq4}, the SPP damping
rate in the presence of the scattering can be written as
\begin{eqnarray}
  \tau^{-1} & = & \sum_{ i} \tau^{-1}_{ i}, \label{eom3:eq2_1}\\
  \tau^{-1}_{ i} & = & \sum_{ q \sigma} \sum_{ (nk, n'k') \in i}
  |p^{ Q_s}_{ q} g^{ nn'}_{ q}|^2 [ f_{ n'\sigma}(k') - f_{
    n\sigma}(k) ] \nonumber\\
  && {} \times \frac{\bar{\Gamma}^b_i}{( \omega^{ nn'}_{ kk'} -
    \bar{\Gamma}^a_i )^2 + (\bar{\Gamma}^b_i)^2} \delta_{
    k'-k-q}. \label{eom3:eq2_2}
\end{eqnarray} 
By comparing the above equations to Eq.~\eqref{eom2:eq4}, one observes that the
$\delta$-function corresponding to the energy conservation Eq.~\eqref{eom:eq_11}
is broadened into a Lorentzian with width $\bar{\Gamma}^{b}_i$ and shift
$\bar{\Gamma}^{a}_i$.

It should be emphasized that the broadening and shifting of the resonance pair
can also be visualized as the broadening and shifting of the corresponding
resonant peak in the polarizations as illustrated in Fig.~\ref{eom:fig2}. This
offers a simply way to interpret the influence of the broadening and shifting
on the SPP damping rate. The influence of the shifting depends
on the direction of the shift. From the figure, one can see that the shifting
reduces the resonant peak in the polarization if the peak is shifted towards the
region with smaller $\delta f$, thus the absorption of the SPP by the
corresponding resonant pairs is reduced [(iii) in Figs.~\ref{eom:fig2}(a) and
(c)] and the corresponding SPP damping rate is suppressed. Otherwise, if the
peak is shifted towards the region with larger $\delta f$ [(iii) in
Fig.~\ref{eom:fig2}(b)], the absorption is enhanced and the SPP damping rate is
enhanced.

The influence of the broadening can be different in the strong and weak Landau
damping regimes at low temperatures as illustrated in Figs.~\ref{eom:fig2}(a)
and (b). In the strong Landau damping regime, the broadening can reduce the
sharp resonance peak of the polarization [(ii) in Fig.~\ref{eom:fig2}(a)] and
suppress the absorption of the SPP. Thus the corresponding SPP damping rate is
suppressed in this regime. In contrast, in the weak Landau damping regime, the
broadening of the resonance increases the absorption from the region with larger
$\delta f$ [(ii) in Fig.~\ref{eom:fig2}(b)], thus the absorption is enhanced and
the corresponding SPP damping rate is enhanced. Note that at high temperatures,
as the polarization exhibits a sharp peak around the resonance in both the
strong and weak Landau damping regimes, the scattering tends to suppress the SPP
damping rate in both regimes [(ii) in Fig.~\ref{eom:fig2}(c)].

\subsection{Broadening and shifting from a simplified model}
\label{sec2_4}

In order to gain a further understanding of the microscopic origin of the
broadening and shifting, we discuss the contributions of the ei, ep,
and ee scatterings to the broadening and shifting within a simplified
model in this section.

Within this model, we assume that the scattering only occurs between the
polarizations inside each resonant pair. Under such assumption, all the
scattering terms can be written in an unified form for the $i$-th resonant pair
\begin{eqnarray}
  \bar{I}^{ i \sigma}_{ nk,n'k'} & = & \sum_q \Big\{ \Gamma^{a}_{nk,n'k',i}(q) P_{\sigma}(nk-q,
  n'k'-q) \nonumber\\
  &&\hspace{0.7cm} {}- \Gamma^{b}_{ nk,n'k',i}(q) P_{\sigma}(nk, n'k') \Big\},
  \label{gr:eq4}
\end{eqnarray}
where both $(nk, n'k')$ and $(nk-q, n'k'-q)$ belong to the $i$-th resonant
pair. Note that the corresponding $\Gamma^{a/b}$ for each scattering mechanism
can be obtained by comparing the scattering terms
[Eqs.~(\ref{eom:eq5}-\ref{eom:eq7})] to $\bar{I}^{ i \sigma}_{ nk,n'k'}$ given
in the above equation.

Note that Eq.~\eqref{gr:eq4} has a similar structure as Eq.~\eqref{eom3:eq0},
thus one can derive the corresponding broadening and shifting following the
similar procedure, yielding
\begin{eqnarray}
  &&\hspace{-1.3cm} \bar{\Gamma}^b_i = \sum_q \Big( \Gamma^b_{ nk,n'k',i}(q) -
  \Gamma^a_{ nk,n'k',i}(q) \frac{ \omega^{ nn'}_{ kk'}}{\omega^{ nn'}_{ k-q, k'-q}} \Big), \label{gr:eq6.1}\\
  &&\hspace{-1.3cm} \bar{\Gamma}^a_i = \pi \sum_q \Gamma^a_{ nk,n'k',i}(q)
  (\omega^{ nn'}_{ kk'} - \omega^{ nn'}_{ k-q, k'-q}) \delta(\omega^{ nn'}_{ k-q, k'-q}). \label{gr:eq6.2}
\end{eqnarray}
The broadening $\bar{\Gamma}^b_i$ and shifting $\bar{\Gamma}^a_i$ due to each
scattering can be evaluated by substituting the corresponding scattering term
into Eqs.~(\ref{gr:eq4}-\ref{gr:eq6.2}). Note that we have omitted the
$k,k'$-dependence of $\bar{\Gamma}^{a(b)}_i$ inside each resonant pair to make
the equation simple and physically transparent. For each
resonant pair, $\bar{\Gamma}^{a(b)}_i$ is evaluated by choosing $(nk, n'k')$
corresponding to the SPP central wave vector $Q_s$ (i.e., $k'-k=Q_s$) since the
line-shape function $p^{Q_s}_q$ is peaked at $q=Q_s$. The shifting
$\bar{\Gamma}^{a}_i$ can be written as
\begin{eqnarray}
  \bar{\Gamma}^{a(\rm{ei})}_i & = & 0,  \label{gr:eq9.1}\\
  \bar{\Gamma}^{a(\rm{ep})}_i & = & 0,  \label{gr:eq11.1}\\
  \bar{\Gamma}^{a(\rm{ee})}_i & = & 2 \pi \sum_{j=1,2} \sum_{\bar{n}} \Pi^{n}_{n'} (\bar{n}\bar{k}_j, 0), \label{gr:eq13.1}
\end{eqnarray}
where
\begin{equation}
  \label{gr:eq13.1.1}
  \Pi^{n}_{n'} (\bar{n}\bar{k}, q) = \sum_{\sigma} \tilde{V}^{ n\bar{n}}_{q}
  \tilde{V}^{ \bar{n} n'}_q f^{>}_{ \bar{n}\sigma}(\bar{k}+q) f^{<}_{
    \bar{n}\sigma}(\bar{k}),
\end{equation}
with $\tilde{V}_q$ being the screened ee
interaction. $\bar{k}_1=-2(\varepsilon^n_{k-Q_s}-\varepsilon^{n'}_{k'}+\Omega_s)/Q_s-Q_s/2$
and
$\bar{k}_2=2(\varepsilon^n_{k}-\varepsilon^{n'}_{k'-Q_s}+\Omega_s)/Q_s+Q_s/2$. $\bar{\Gamma}^{a(\rm{ei})}_i$,
$\bar{\Gamma}^{a(\rm{ep})}_i$ and $\bar{\Gamma}^{a(\rm{ee})}_i$ correspond to
the contributions from the ei, ep and ee scatterings, respectively. The
corresponding broadenings have the forms
\begin{eqnarray}
  \bar{\Gamma}^{b(\rm{ei})}_i & = & m^{\ast} \pi n_i ( \frac{\tilde{v}^{ nn}_0}{|k|}
  - \frac{\tilde{v}^{ n'n'}_0}{|k'|} ) ( \tilde{v}^{ nn}_0 -
  \tilde{v}^{ n'n'}_0 ), \label{gr:eq9.2} \\
  \bar{\Gamma}^{b(\rm{ep})}_i & = & \sum_{\sigma} \Big\{ \pi m^{\ast} ( \sum_{\bm{Q}} M^{
    nn}_{\bm{Q},-q} M^{ nn'}_{\bm{Q},-q} ) \nonumber\\
  &&\hspace{-0.8cm}{} \times \Big. [ N^{>}_{\rm LO} f^{>}_{n\sigma}(k-q) +
  N^{<}_{\rm LO} f^{<}_{n\sigma}(k-q) ]/|k-q| \Big|_{ q=q^{(k)}_{+}} \nonumber\\
  &&\hspace{-0.8cm} \mbox{}+ \pi m^{\ast} ( \sum_{\bm{Q}} M^{ nn}_{\bm{Q},-q} M^{ nn'}_{\bm{Q},-q} ) \nonumber\\
  &&\hspace{-0.8cm}{} \times \Big. [ N^{<}_{\rm LO} f^{>}_{n\sigma}(k-q) +
  N^{>}_{\rm LO} f^{<}_{n\sigma}(k-q) ]/|k-q| \Big|_{ q=q^{ (k)}_{-}} \Big\}
  \nonumber\\
  &&\hspace{-0.4cm} {}+ \{ nk \longleftrightarrow n'k' \}, \label{gr:eq11.2} \\
  \bar{\Gamma}^{b(\rm{ee})}_i & = & \Big\{ 2 \pi \sum_{ \bar{n} \bar{k}}
  [ \Pi^{n}_{n} (\bar{n}\bar{k}, 0) - \Pi^{n}_{n'} (\bar{n}\bar{k},
  0) ]/|k-\bar{k}| \Big\} \nonumber\\
  &&\hspace{-0.4cm}{}+ \{ nk \longleftrightarrow n'k' \}. \label{gr:eq13.2}
\end{eqnarray}
where $q^{ (k)}_{\pm}$ satisfies $[q^{ (k)}_{\pm}]^2 - 2 k q_{\pm} \pm 2
m^{\ast} \Omega_{\rm LO} = 0$ and $\tilde{v}^{nn'}_q$ represents the screened ei
interaction. $\{ nk \longleftrightarrow n'k' \}$ stands for the same term as in
the previous $\{\}$ but with the interchange of indices $nk \longleftrightarrow
n'k'$.

From Eqs.~(\ref{gr:eq9.1}-\ref{gr:eq13.2}), one finds that different
scattering has different contribution
 to the broadening and shifting. Only the ee scattering
contributes to the shifting, while the contributions from the ei and ep
scattering vanish. Although all the scatterings can contribution to the
broadening, their relative importance can be quite different. This is because
for the nanowires considered here, the ei, ep and ee matrix elements $v^{
  nn'}_q$, $M^{ nn'}_{\bm{Q} q}$ and $V^{nn'}_q$ are not sensitive to the
subband index $n$. Thus, for the ei and ee scatterings, according to
Eqs.~\eqref{gr:eq9.2} and~\eqref{gr:eq13.2}, the terms in the bracket can
largely cancel each other, leading to small contributions left. In contrast,
such cancellation is absent for the ep scattering, thus its contribution to the
broadening is expected to be larger than the ones from the ei and ee
scatterings.

Equations~\eqref{eom3:eq2_1} and~\eqref{eom3:eq2_2} with the broadening and
shifting given in Eqs.~(\ref{gr:eq9.1}-\ref{gr:eq13.2}) consist the analytic
solution for the SPP damping rate. The analytic solution in this section
offers a simple picture to understand the influence of the scattering on the SPP
damping: The SPP damping comes from the resonant absorption of the SPP by
electrons, while the scattering can introduce a broadening and a shifting to the
resonance and hence affects the damping process. At low temperatures, the
broadening tends to suppress the SPP damping rate in the strong Landau damping
regime. While in the weak Landau damping regime, the broadening tends to enhance
the SPP damping. At high temperatures, such difference vanishes and the
broadening tends to suppress the damping in both regimes. The shifting can
suppress the SPP damping if the resonance is shifted towards the region with
smaller $\delta f$, but boost it if the resonance is shifted toward the region
with larger $\delta f$. Different scatterings can have different contributions
to the broadening and shifting. From the simplified model in this section,
one finds that the shifting is determined by the ee scattering, and the
broadening is mainly decided by the ep scattering for the typical
nanowires considered here.
 
\section{Numerical Results}
\label{sec5}

In the numerical investigation, we choose nanowires to be free-standing InAs
nanowires. The typical electron density $\bar{n}_0$ is in the range of $10^{17}
\sim 10^{18}$~cm$^{-3}$ and the wire radius $R$ is around $25 \sim
75$~nm.\cite{Dayeh2007} The LO phonon energy $\Omega_{\rm LO} = 29$~meV and the
electron effective mass $m^{\ast}=0.023 m_0$ with $m_0$ representing the free
electron mass. The dielectric constants of the nanowires are
$\epsilon^{\infty}_1=12.3$ for high frequency and $\epsilon^0_1=15.5$ for low
frequency. The dielectric constant outside the nanowire is
$\epsilon_2=1.0$. Note that for such nanowires, there are $10$-$20$ electron
subbands relevant to the SPP damping. We set the impurity line density $n_i=0.5
n_e$ with $n_e=\pi R^2 \bar{n}_0$ being the electron line density.

By numerical solving the kinetic equations Eqs.~\eqref{eom:eq4_1}
and~\eqref{eom:eq4_2}, one obtains the temporal evolution of the SPP amplitude
$\bar{B}_s$. The SPP damping rate $\tau^{-1}$ can be extracted by fitting the
real part of $\bar{B}_s$ with a single exponential decay of the cosine
oscillation: $\text{Re}[\bar{B}_s] = \bar{B}^0_s \exp(-t/\tau) \cos(\omega_s
t)$, where the initial value of the SPP amplitude $\bar{B}^0_s$ is chosen to be
real. We set the wave packet length $L=100 R$
  (Ref.~\onlinecite{comment2}).

\subsection{Landau damping: Size and temperature dependence}
\label{sec5_1}

\begin{figure*}
  \centering
  \includegraphics[width=7.5cm]{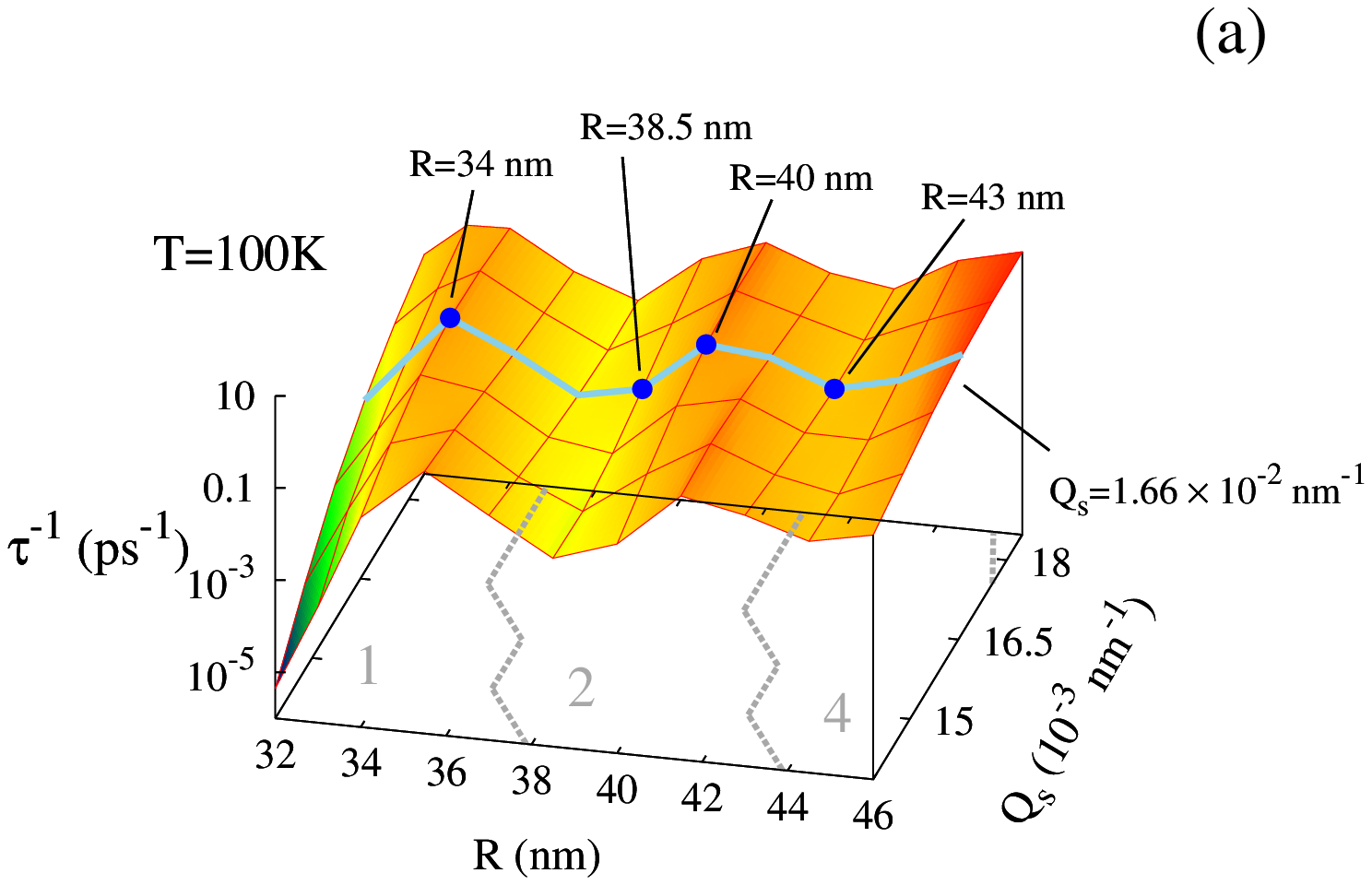}
  \includegraphics[width=7.0cm]{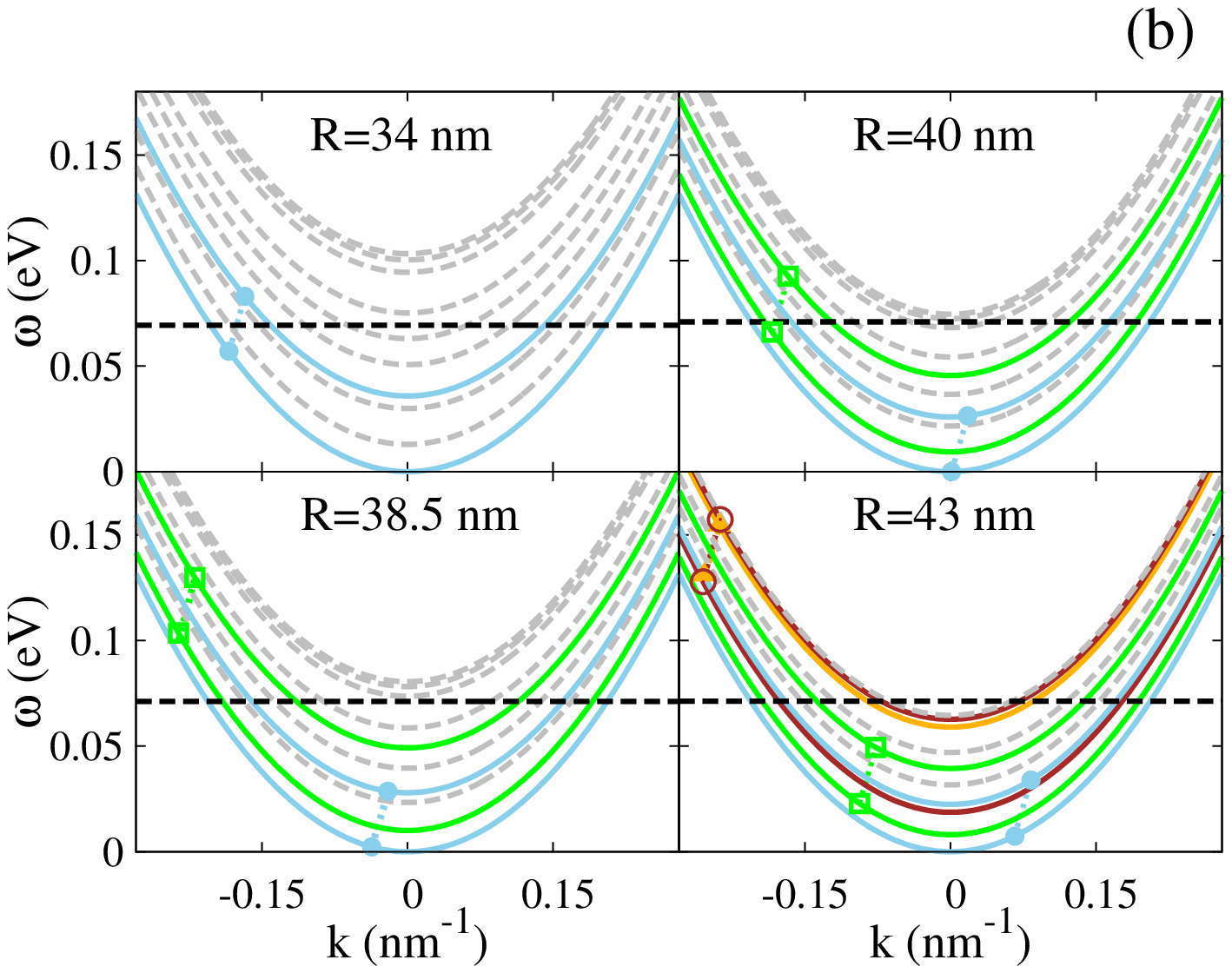}
  \includegraphics[width=7.5cm]{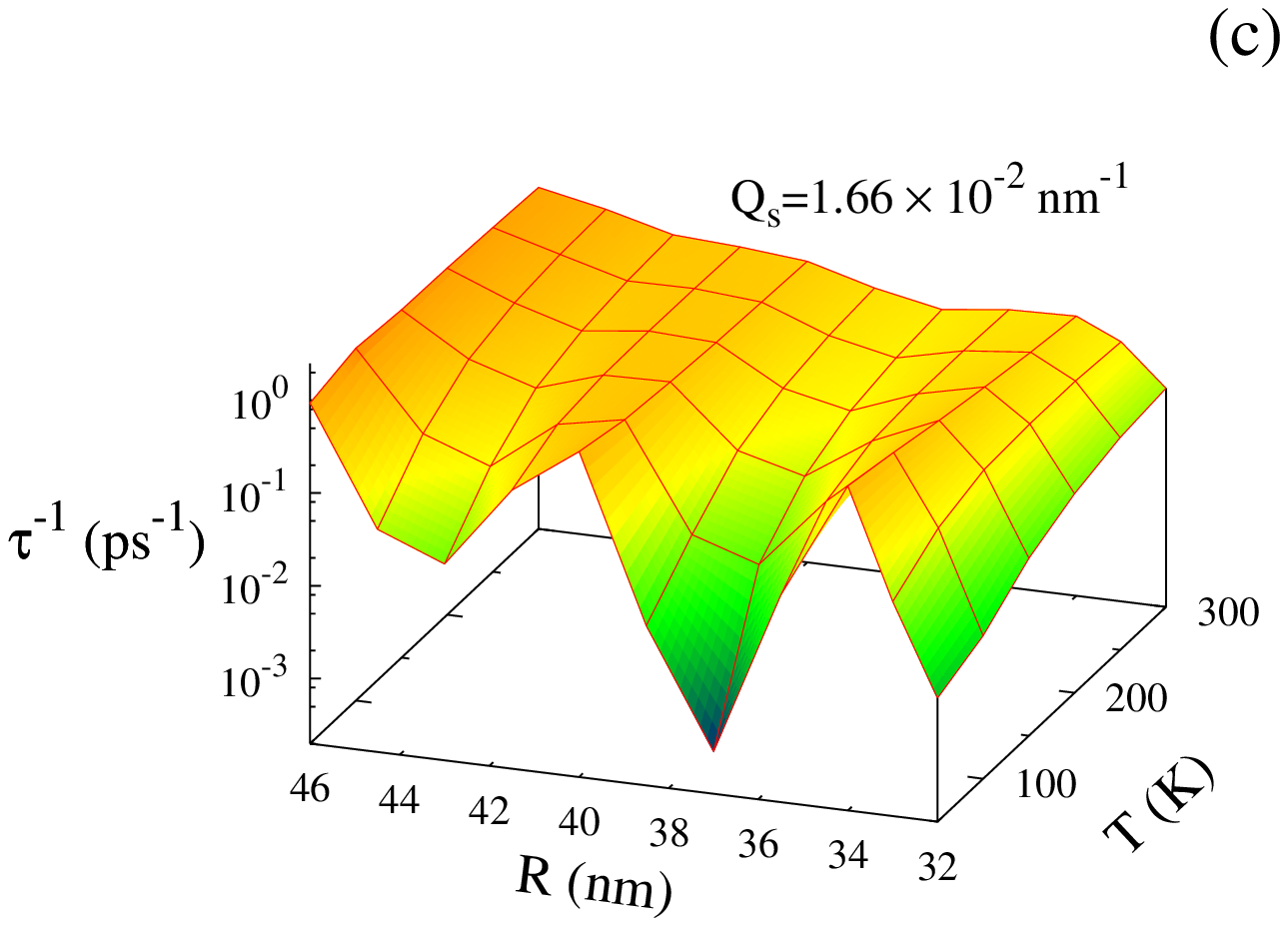}
  \includegraphics[width=7.0cm]{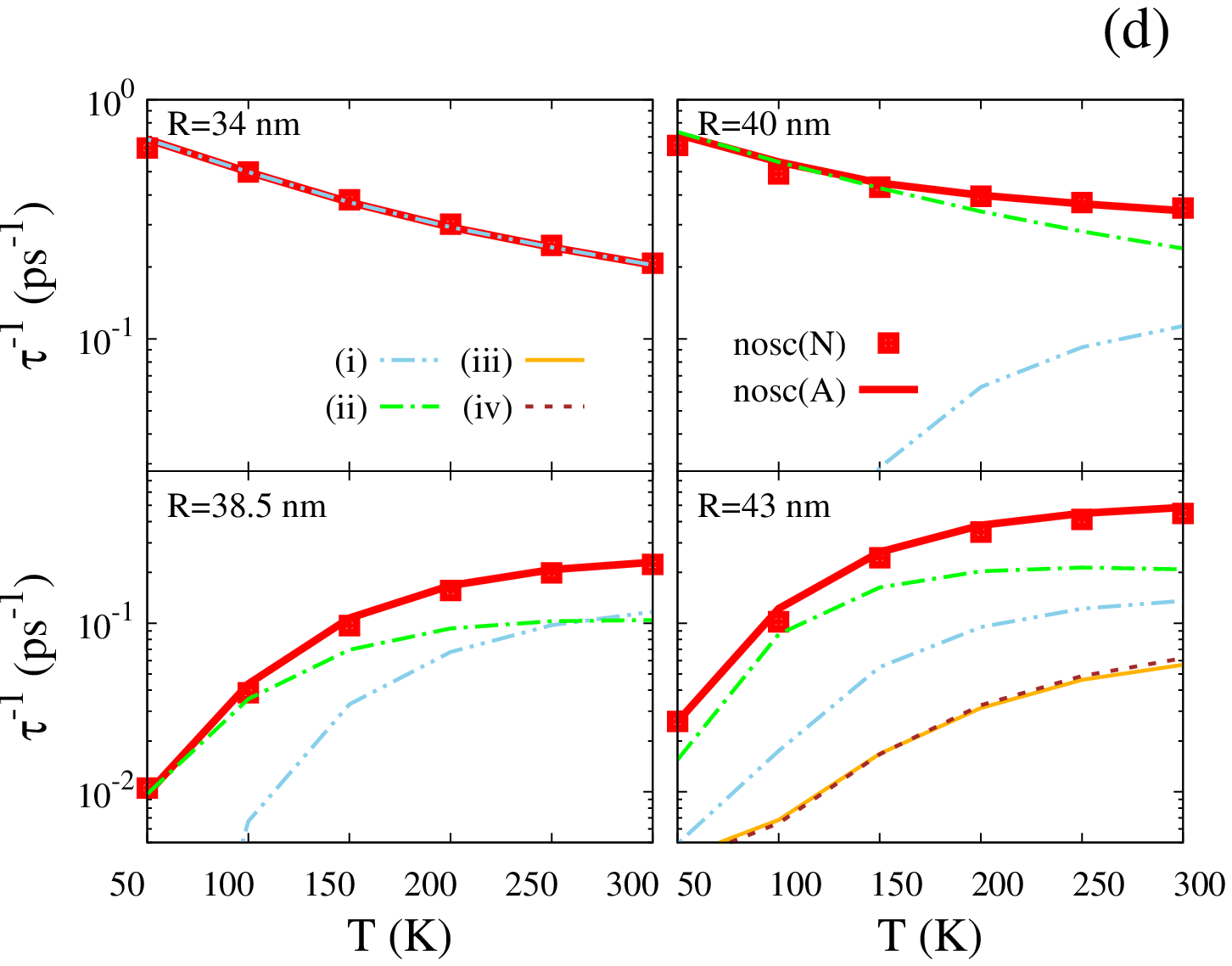}
  \caption{(Color online) (a) The SPP damping rate as function of SPP central
    wave vector $Q_s$ and wire radius $R$ without scattering for $\bar{n}_0=1.5
    \times 10^{17}$~cm$^{-3}$ and $T=100$~K. The skyblue curve represents the
    damping rate corresponding to the SPP central wave vector $Q_s=1.66 \times
    10^{-2}$~nm$^{-1}$. The number of the relevant resonant pairs are labelled
    in the $R$-$Q_s$ plane. (b) The resonant pairs for $R=34$, $38.5$, $40$ and
    $43$~nm corresponding to $Q_s=1.66 \times 10^{-2}$~nm$^{-1}$ in the electron
    spectrum. For clarification, only the lowest $10$ subbands are plotted. (c)
    The SPP damping rate without scattering as function of radius $R$ and
    temperature $T$ for $Q_s=1.66 \times 10^{-2}$~nm$^{-1}$. (d) Temperature
    dependence of the damping rate $\tau^{-1}$ without scattering for $R=34$,
    $38.5$, $40$ and $43$~nm with $Q_s=1.66 \times 10^{-2}$~nm$^{-1}$. Symbols
    correspond to the numerical results. Red curves represent the results from
    the analytic solution. The contributions of each resonant pair from the
    analytic solution are also plotted as curves with different colors. The
    skyblue double-dotted chain, green chain, brown dotted and yellow solid
    curves correspond to the contribution from the resonant pair (i-iv),
    respectively.}
  \label{re:fig0}
\end{figure*}

Before we discuss the influence of the scattering, it is helpful to first obtain
an understanding of the SPP damping without scattering. In
Fig.~\ref{re:fig0}(a), we show a typical behavior of the SPP damping rate as
function of the SPP central wavevector $Q_s$ and wire radius $R$ for the
nanowire with electron density $\bar{n}_0=1.5 \times 10^{17}$~cm$^{-3}$. The
temperature is chosen to be $100$~K. One finds that the SPP damping rate
oscillates with the radius $R$. Note that similar size-dependent oscillations
have also been reported in metal nanoparticles and thin films.\cite{Charle1989,
  Brechignac1993, Mochizuki1997, Molina2002, Weick2005, Gao2011}

Such oscillations are usually attributed to the quantized electron states in the
nanostructures,\cite{Molina2002, Weick2005, Gao2011} which can be understood in
terms of the resonant pairs in the nanowires we studied here. To illustrate
this, we concentrate on the damping rate corresponding to a typical SPP central
wavevector $Q_s=1.66 \times 10^{-2}$~nm$^{-1}$ [skyblue curves in
Fig.~\ref{re:fig0}(a)] and show the corresponding resonant pairs for radii
$R=34$, $38.5$, $40$ and $43$~nm in Fig.~\ref{re:fig0}(b). Each resonant pair
can be represented by the resonance corresponding to the central wave vector
$Q_s$, since the line-shape function of the SPP wave packet is peaked at
$Q_s$. There are four resonant pairs in Fig.~\ref{re:fig0}(b) which lay between
different subbands: pair (i) is between the subbands $1$ and $4$, pair (ii) is
between the subbands $2$ and $6$, pair (iii) is between the subbands $3$ and $9$
and pair (iv) is between the subbands $4$ and $8$. In the figure, the four
resonant pairs (i-iv) are denoted by the skyblue dots, green squares, brown open
circles and yellow triangles, respectively. The subbands corresponding to each
resonant pair are also plotted with solid curves in the same color. Note that
for $R=34$~nm, only the resonant pair (i) is relevant for the damping. For
$R=38.5$ and $40$~nm, both the resonant pairs (i) and (ii) are relevant. For
$R=43$~nm, all the four resonant pairs (i-iv) contribute to the SPP damping.

From Fig.~\ref{re:fig0}(b), one can see that as the radius $R$ increases, the
resonant pairs move from left to right in the electron spectrum. When a resonant
pair moves across the chemical potential marked by the horizontal black dashed
lines, a crossover between the strong and weak Landau damping regimes
occurs, which induces the size-dependent oscillations
shown in Fig.~\ref{re:fig0}(a). Note that the peaks/valleys correspond to the
strong/weak Landau damping regimes. For example, the oscillation from $R=32$ to
$38.5$~nm is due to the crossover induced by the resonant pair (i). While the
crossover induced by the resonant pair (ii) induces the oscillation from
$R=38.5$ to $43$~nm. One finds from the figure that $R=34$ and $40$~nm
correspond to the strong Landau damping regime whereas $R=38.5$ and $43$~nm
correspond to the weak Landau damping regime.

One also observes from Fig.~\ref{re:fig0}(b) that due to the many subbands in
the nanowires, there usually exist multi-resonant pairs relevant for the SPP
damping. For a given $Q_s$, more and more resonant pairs become involved as the
radius $R$ increases. The number of relevant resonant pairs are labeled in the
$R$-$Q_s$ plane in Fig.~\ref{re:fig0}(a). Note that as the number of the
resonant pairs increases, the magnitude of the size-oscillations become less
pronounced. This is mainly because the oscillations are usually induced by the
crossover due to one resonant pair as $R$ varies. For the system with many
resonant pairs, the contribution from one resonant pair becomes less
significant. Thus the size-dependent oscillations can be suppressed for
nanowires with large $R$.

It should be emphasized that the size-dependent oscillations can also be
suppressed by increasing temperature $T$. This is because the crossover is more
pronounced for strongly degenerate electrons where a clear Fermi surface exists
around the chemical potential. In high-temperature regime, the crossover is
largely suppressed. To show this, we plot the damping rate $\tau^{-1}$ as
function of the radius $R$ and temperature $T$ for $Q_s=1.66 \times
10^{-2}$~nm$^{-1}$ in Fig.~\ref{re:fig0}(c). One sees that as temperature
increases, the damping rate $\tau^{-1}$ corresponding to the strong Landau
damping regime decreases, while $\tau^{-1}$ corresponding to the weak Landau
damping regime increases. This leads to a suppression of the size-dependent
oscillations in  high-temperature regime.

One can also obtain the above results from the analytic solution of the kinetic
equations without scattering [Eqs.~\eqref{eom2:eq5_1}
and~\eqref{eom2:eq5_2}]. To show this, we compare the temperature dependence of
$\tau^{-1}$ from both the numerical (red squares) and analytic (red solid
curves) solutions for the nanowires with $R=34$, $38.5$, $40$ and $43$~nm in
Fig.~\ref{re:fig0}(d). One finds good agreement between each other, indicating
that the analytic solution without scattering offers a good estimation to the
numerical results. Note that according to the analytic solution, the temperature
dependence of the SPP damping rate originates from the population difference of
the resonant pairs.

From the analytic solution, one can also identify contributions from different
resonant pairs, which are plotted as curves with different colors and line
shapes in Fig.~\ref{re:fig0}(d). The skyblue double-dotted chain, green chain,
brown dotted and yellow solid curves correspond to the contributions from the
resonant pair (i-iv), respectively. It is clear that the relative importance of
the resonant pairs can be quite different. For the strong Landau damping regime,
there usually exists one resonant pair whose contribution is much larger than
the other pairs. For example, the damping rate $\tau^{-1}$ is mainly determined
by the resonant pairs (i) and (ii) for $R=34$~nm and $R=40$~nm,
respectively. For the weak Landau damping regime, the contributions from
different resonant pairs can be comparable. For example, for $R=38.5$ and
$R=43$~nm, although the resonant pair (ii) has a large contribution to the
damping rate $\tau^{-1}$, the other resonant pairs can also play important
roles, especially at high temperatures.

From the above results, one finds that the SPP damping exhibits size-dependent
oscillations and distinct temperature dependence without scattering, which can
be explained by the analytic solution.

\subsection{Influence of scattering}
\label{sec5_2}

\begin{figure}
  \centering
  \includegraphics[width=7.5cm]{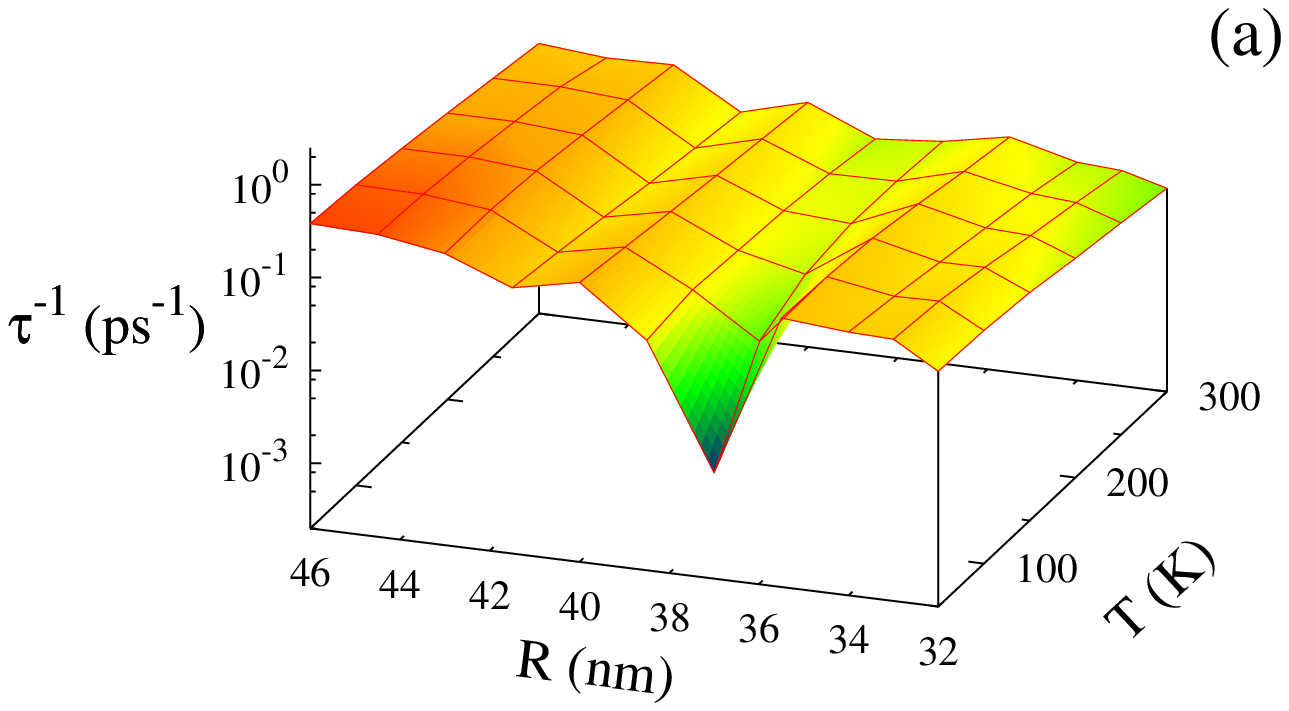}
  \includegraphics[width=7.5cm]{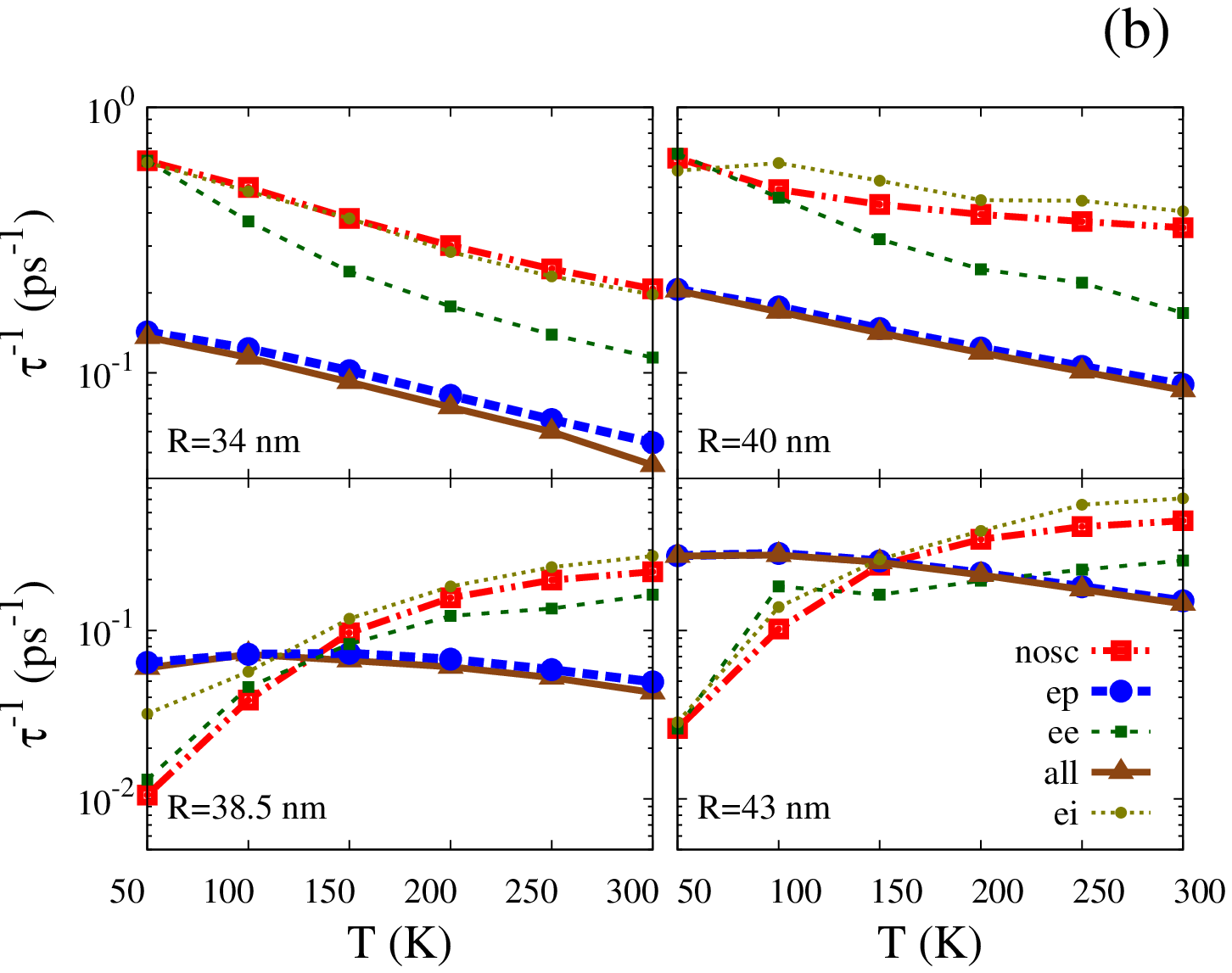}
  \caption{(Color online) (a) SPP damping rate in the presence of all the
    scattering as function of radius $R$ and temperature $T$ for $Q_s=1.66
    \times 10^{-2}$~nm$^{-1}$. (b) SPP damping rate $\tau^{-1}$ calculated from
    the numerical results for $Q_s=1.66 \times 10^{-2}$~nm$^{-1}$ with different
    scattering for $R=34$, $38.5$, $40$ and $43$~nm. Symbols with big blue dots,
    small green squares, small olive dots represent $\tau^{-1}$ calculated with
    the ep, ee and ei scatterings, respectively. The brown triangles represent
    $\tau^{-1}$ calculated in the presence of all the three scatterings and the
    red squares represent $\tau^{-1}$ without scattering.}
  \label{re:fig2}
\end{figure}

Now we discuss the influence of the scattering on the SPP damping. In
Fig.~\ref{re:fig2}(a), we plot the damping rate $\tau^{-1}$ as function of the
radius $R$ and temperature $T$ for $Q_s=1.66 \times 10^{-2}$~nm$^{-1}$ in the
presence of all the scattering. Comparing to the case without scattering
[Fig.~\ref{re:fig0}(c)], one finds that the scattering has pronounced influence
on the SPP damping: (1) The size-dependent oscillations are effectively smeared
out, and (2) the temperature dependence also becomes weaker compared to the case
without scattering.

To gain a better understanding of the influence of the scattering, in
Fig.~\ref{re:fig2}(b), we compare the temperature dependence of the damping rate
$\tau^{-1}$ with and without scattering for nanowires with four typical radii
$R=34$, $38.5$, $40$ and $43$~nm. The brown triangles represent $\tau^{-1}$
calculated in the presence of all the three scatterings, while $\tau^{-1}$
without scattering are plotted with red squares for comparison. Note that $R=34$
and $40$~nm correspond to the strong Landau damping regime, whereas $R=38.5$ and
$43$~nm correspond to the weak one.

In the presence of the scattering, it is seen that the damping rate $\tau^{-1}$
is markedly suppressed in the strong Landau damping regime ($R=34$ and
$40$~nm). In contrast, for the weak Landau damping regime ($R=38.5$ and
$43$~nm), the scattering plays different roles in different temperature regimes:
The damping rate is markedly enhanced in the low-temperature regime ($T\lesssim
150$~K), but is largely suppressed in the high-temperature regime ($T\gtrsim
150$~K). A crossover exists at the intermediate temperature regime.  It is also
noted that the damping rate can be enhanced/suppressed by almost one order of
magnitude by the scattering.

\begin{figure}
  \centering
  \includegraphics[width=7.5cm]{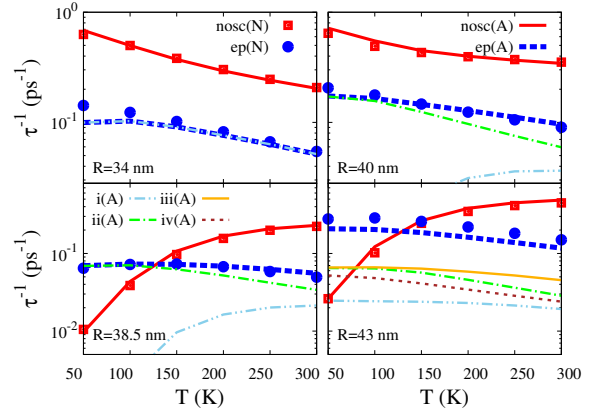}
  \caption{(Color online) Comparison between the numerical and analytical
    results of the SPP damping rate for $R=34$, $38.5$, $40$ and $43$~nm with
    the ep scattering. Blue dots represent the numerical results, while the blue
    dashed curves show the results from the analytic solution. The contribution
    of each resonant pair from the analytic solution is also plotted. The
    skyblue double-dotted chain, green chain, brown dotted and yellow solid
    curves correspond to the contribution from the resonant pair (i-iv) ,
    respectively. For comparison, the numerical and analytical results for the
    damping rate without scattering are also plotted with red squares and solid
    curves, respectively.}
  \label{re:fig1}
\end{figure}

To understand these influences, we first identify the dominant scattering
mechanism. To do so, we calculate the damping rates $\tau^{-1}$ with the ep, ee
or ei scattering only, and plot them with big blue dots, small green squares and
small olive dots in Fig.~\ref{re:fig2}(b), respectively. It is clear to see from
the figure that the damping rate $\tau^{-1}$ is dominated by the ep
scattering.\cite{comment3}

We first concentrate on the ep scattering. From the analytic solution within the
simplified model, we have attributed the effect of the ep scattering to the
broadening of the resonant pairs. To see if this picture gives a proper
description of the influence of the ep scattering in general case, we calculate
the temperature dependence of the damping rate $\tau^{-1}$ by using the analytic
solution Eqs.~\eqref{eom3:eq2_1} and~\eqref{eom3:eq2_2} with the
ep-scattering--induced broadening given in Eq.~\eqref{gr:eq11.2}. The calculated
analytic results are compared to the numerical ones in Fig.~\ref{re:fig1}.

In the figure, the blue dots represent the damping rate $\tau^{-1}$ from the
numerical results, while the blue dashed curves represent $\tau^{-1}$ from the
analytic results. For comparison, we also plot the numerical and analytic
results without scattering with red squares and solid curves, respectively. One
finds good agreement between each other, indicating that the broadening can give
a proper description of the effect of the ep scattering. Note that at low
temperatures, the broadening tends to suppress the SPP damping rate in the
strong Landau damping regime ($R=34$ and $40$~nm). While in the weak Landau
damping regime ($R=38.5$ and $43$~nm), the broadening tends to enhance the SPP
damping rate. At high temperatures, such difference vanishes and the SPP damping
rate is suppressed in both the strong and weak Landau damping regimes. This
leads to a crossover between the suppression and enhancement for $R=38.5$ and
$43$~nm corresponding to the weak Landau damping regime as shown in
Fig.~\ref{re:fig1}. This also agrees with the conclusion from the analytic
solution.

It is worth noting that the scattering can also suppress the temperature
dependence of the SPP damping rate. This is because the temperature dependence
originates from the population difference of the resonant pairs. For the
resonant pairs with the broadening, the corresponding population difference is
less sensitive to the temperature, leading to the suppression of the temperature
dependence of the corresponding SPP damping rate.

We further point out that the scattering can also change the relative importance
of different resonant pairs. To see this, we identify contributions of different
resonant pairs from the analytic solution, as applied in Sec.~\ref{sec5_1} for
the case without scattering. In Fig.~\ref{re:fig1}, the skyblue double-dotted
chain, green chain, brown dotted and yellow solid curves correspond to the
contributions from the resonant pairs (i-iv), respectively. One finds that the
resonant pair (i) dominates the damping for $R=34$~nm, whereas the resonant pair
(ii) plays the most important role for $R=38.5$ and $40$~nm. For $R=43$~nm, all
the four resonant pairs have comparable contributions to the damping, with the
largest contribution coming from the resonant pair (iii). Comparing to the case
without scattering [Fig.~\ref{re:fig0}(d)], one also finds that the relative
importance of different resonant pairs is modified by the broadening, especially
in weak Landau damping regime.

\begin{figure}
  \centering
  \includegraphics[width=7.5cm]{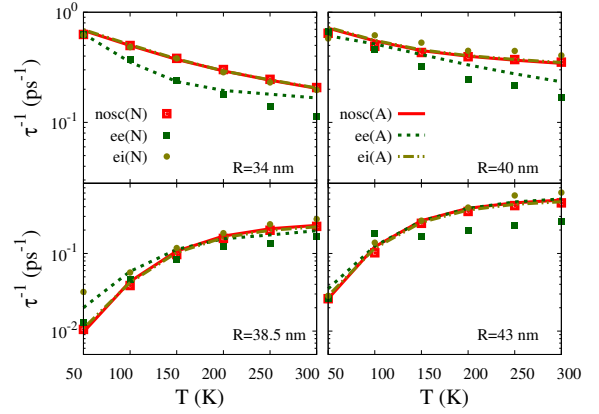}
  \caption{(Color online) Comparison between the numerical and analytical
    results of the SPP damping rate for the ee and ei scatterings for $R=34$,
    $38.5$, $40$ and $43$~nm. Small green squares and green dotted curves
    represent the numerical and analytical results for the ee scattering,
    respectively. Small olive dots and olive double-dotted chain represent
    the numerical and analytical results for the ei scattering,
    respectively. For comparison, the numerical and analytical results for the
    damping rate without scattering are also plotted with red squares and solid
    curves, respectively.}
  \label{re:fig3}
\end{figure}

From the above discussion, one comes to the conclusion that the influence of the
ep scattering on the SPP damping rate can be understood as the broadening of the
resonant pairs. The analytic solution incorporating such broadening shows good
agreement with the numerical result. Note that in the above results, the
contribution of the SO phonons is omitted since it is much smaller than that
from the LO phonons. This is shown in detail in Appendix~\ref{app5}.

Now we briefly address the ee and ei scatterings. From the analytic solution
within the simplified model, the effect of the ei scattering is attributed to
the broadening of the resonant pairs. While for the ee scattering, the main
effect is due to the shifting. The SPP damping rate with the ee/ei scattering
can also be calculated from the analytic solution [Eqs.~\eqref{eom3:eq2_1}
and~\eqref{eom3:eq2_2} with the broadening and shifting given in
Eqs.~\eqref{gr:eq9.1} and~\eqref{gr:eq9.2} for the ei scattering and
Eqs.~\eqref{gr:eq13.1} and~\eqref{gr:eq13.2} for the ee scattering,
respectively]. The analytic results are compared to the numerical ones in
Fig.~\ref{re:fig3}. From the figure, one observes qualitatively good agreement
between each other, indicating that the ee and ei scattering can also be
understood as the broadening and/or shifting of the resonant
pairs.

It is pointed out that the effect of the broadening and shifting can be
visualized from the behavior of the polarization, which gives a more intuitive
picture for the effect of the scattering. This is discussed in detail in
Appendix~\ref{app4}.

\begin{figure}
  \centering
  \includegraphics[width=7.5cm]{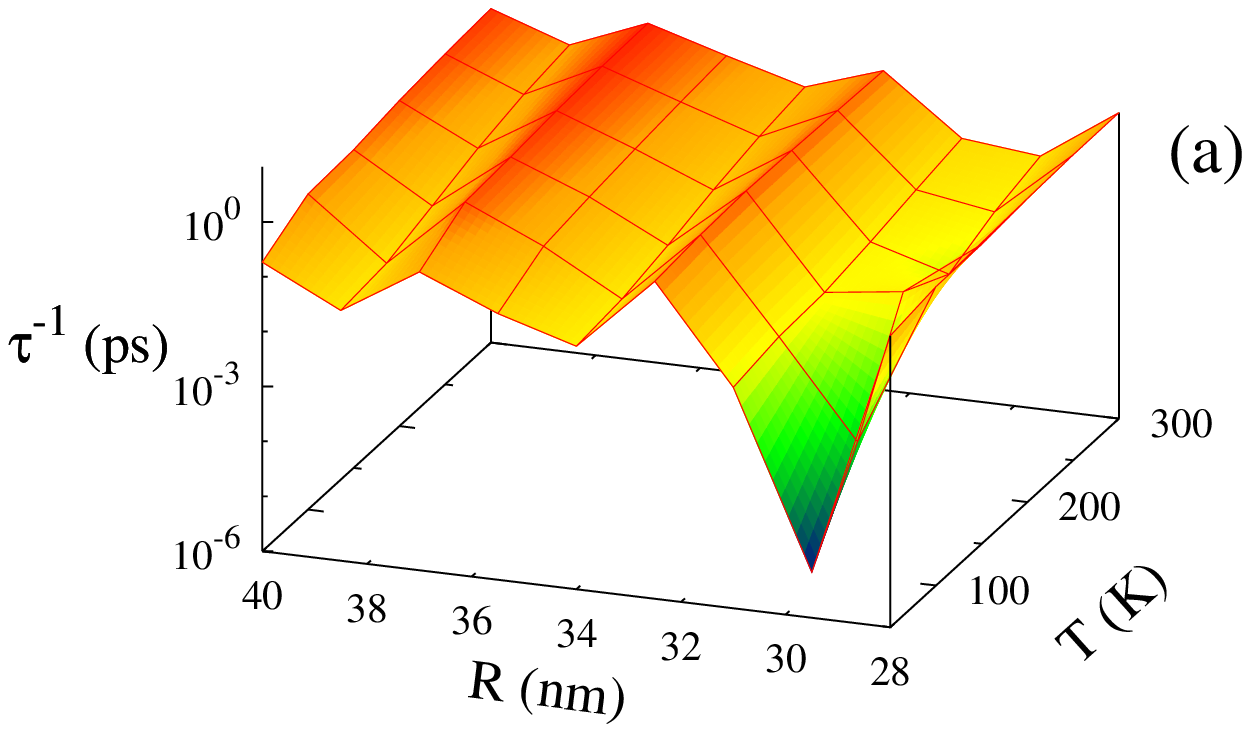}
  \includegraphics[width=7.5cm]{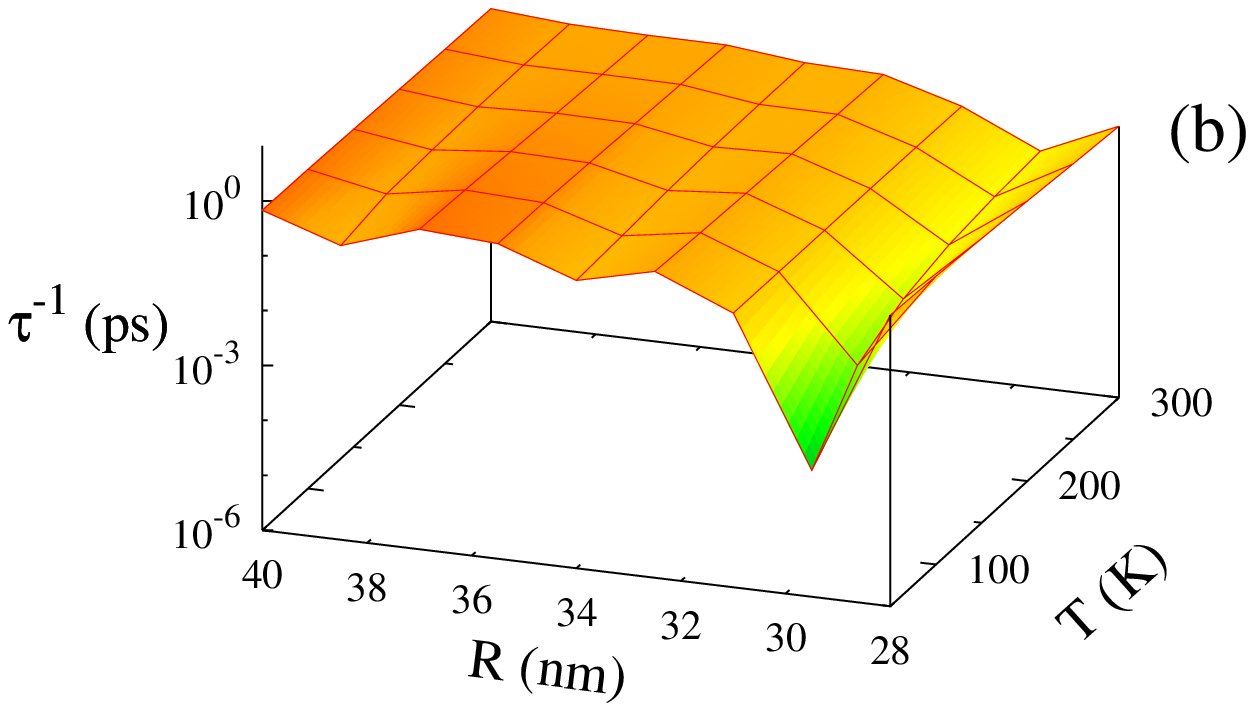}
  \caption{(Color online) Size and temperature dependence of the SPP damping
    rate for nanowires with electron density $\bar{n}_0=5.0 \times
    10^{17}$~cm$^{-3}$ (a) without any scattering and (b) with all the
    scattering. The central wavevector of the SPP wave packet is chosen to be
    $Q_s=1.3 \times 10^{-2}$~nm$^{-1}$.}
  \label{re:fig6}
\end{figure}

From the above results, one finds that the scattering tends to smear out the
size-dependent oscillations of the SPP damping rate. The temperature dependence
can also be suppressed by the scattering. Note that this effect is quite general
and can be seen for nanowires with different electron densities. To demonstrate
this, we show the size and temperature dependence of the SPP damping rates for
nanowires with electron density $\bar{n}_0=5.0 \times 10^{17}$~cm$^{-3}$ without
and with scattering in Figs.~\ref{re:fig6}(a) and (b), respectively. The central
wavevector of the SPP wave packet is chosen to be $Q_s=1.3 \times
10^{-2}$~nm$^{-1}$. From the figure, it is seen that the size-dependent
oscillations are smeared out by the scattering. The temperature dependence is
also suppressed. These effects are similar to the ones for nanowires with
$\bar{n}_0=1.5 \times 10^{17}$~cm$^{-3}$ investigated above, indicating that
these effects are quite general for typical InAs nanowires.

\section{Conclusion and discussion}
\label{sec8}

In conclusion, we present a microscopic kinetic theory to study the effect of
the electron scattering on the Landau damping of the SPP in semiconductor
nanowires. Based on the semiclassical model Hamiltonian of the SPP-electron
system and the nonequilibrium Green-function approach, we derive the kinetic
equations of the SPP-electron system, with all the scattering explicitly
included. Within this model, the SPP damping is understood as the absorption of
the SPP by the electron polarization of the resonant pairs. The population
difference of the resonant pairs $\delta f$ plays an important role on the SPP
damping, leading to a strong and a weak Landau damping regimes for degenerate
electrons. The scattering influences the SPP damping via the broadening and
shifting of the resonant pairs, which have different effects on the strong and
weak Landau damping regimes. At low temperatures, the broadening tends to
suppress the SPP damping in the strong Landau damping regime. Whereas in the
weak Landau damping regime, the broadening tends to enhance the SPP damping. At
high temperatures, this difference tends to be vanished. The shifting can
suppress the SPP damping if the resonance is shifted towards the region with
smaller $\delta f$, but boost it if the resonance is shifted toward the region
with larger $\delta f$. The broadening and shifting can be visualized from the
corresponding electron polarization of the resonant pairs. Moreover, different
scattering has different contribution to the broadening and shifting. The main
effect of the ei/ep scattering is to cause a broadening, whereas the main effect
of the ee scattering is to introduce a shifting. The effect of the broadening
and shifting can be incorporated into an analytic solution, which shows good
agreement with the numerical result.

To demonstrate the effect of the scattering, we investigate the damping of the
axial symmetric SPP mode in InAs nanowires in the presence of the ei, ee and ep
scatterings. Without any scattering, the SPP damping exhibits size-dependent
oscillations and a distinct temperature dependence. In the presence of the
scattering, the size-dependent oscillations are markedly smeared out and the
temperature dependence is also suppressed. The damping rate can be
  enhanced/suppressed by almost one order of magnitude. For InAs nanowires
investigated here, the ep scattering is found to be dominant. These results are
found to be general for typical InAs nanowires, which demonstrate the importance
of the scattering on the SPP damping for semiconductor nanowires. It is further
pointed out that our model can be applied to nanowires made of other
semiconductors, offering a systematic way to investigate the effect of
electron scattering on the SPP damping in such systems.

\begin{acknowledgments}
  This work was supported by the National Basic Research Program of China under
  Grant No. 2012CB922002 and the Strategic Priority Research Program of the
  Chinese Academy of Sciences under Grant No. XDB01000000. One of the authors
  (YY) was also partially supported by the China Postdoctoral Science
  Foundation.
\end{acknowledgments}

\appendix

\section{Semiclassical model for SPP}
\label{app1}

Our derivation of the quantized SPP Hamiltonian $H_{\rm SPP}$ and its coupling
to electrons $H_{\rm SPP\text{-}el}$ follows the procedure used in
Ref.~\onlinecite{Bergara1999}, in which the SPPs are obtained from quantization
of the classical plasmon field. The classical plasmon field is derived within
the hydrodynamical model, which treats plasmons as irrotational deformations of
the conduction electrons.\cite{Ritchie1969, Barton1979, Bergara1999, Perez2009,
  Prodan2003} From this model, the Hamiltonian of the plasmon can be written
as\cite{Bergara1999}
\begin{eqnarray} 
  H_{\rm SPP} & = & \int d\bm{r} \frac{m^{\ast}}{2} n(\bm{r}) | \bm{\nabla}
  \eta(\bm{r}) |^2 \nonumber\\
  &&\mbox{}+ \frac{e^2}{2} \int d\bm{r} \frac{e^2}{2} \int d\bm{r}'
  n(\bm{r}) n(\bm{r}') V_{\rm ee}(\bm{r}, \bm{r}') \nonumber\\
  &&\mbox{}+ \int d\bm{r} G[ n(\bm{r}) ],
  \label{app1:eq1}
\end{eqnarray}
where the irrotational flow has been assumed, i.e., $\bm{v}(\bm{r}) =
\bm{\nabla} \eta(\bm{r})$ with $\bm{v}(\bm{r})$ being the velocity
field. $n(\bm{r})$ is the electron density and $m^{\ast}$ represents the
electron effective mass. 

For both $\bm{r}$ and $\bm{r}'$ inside the nanowire, the Coulomb interaction
$V_{\rm ee}(\bm{r}, \bm{r}')$ has the form\cite{Slachmuylders2006}
\begin{eqnarray}
  \label{app1:eq3_1}
  V_{\rm ee}(\bm{r}, \bm{r}') & = & \frac{e}{4 \pi \epsilon^{\infty}_1}\Big[
  \frac{1}{|\bm{r} -\bm{r}'|} + \frac{2}{\pi}(\frac{\epsilon^{\infty}_1}{\epsilon_2}-1)
  \sum^{+\infty}_{m=-\infty} e^{i m (\varphi - \varphi')} \nonumber\\
  &&\hspace{-2cm} \times \int^{\infty}_0 dk \cos[ k(z-z') ] \mathbb{C}_m(kR,
  \frac{\epsilon^{\infty}_1}{\epsilon_2}) I_m(k \rho) I_m(k \rho') \Big],
\end{eqnarray}
where
\begin{equation}
  \label{app1:eq4}
  \mathbb{C}_m(kR,\frac{\epsilon^{\infty}_1}{\epsilon_2}) =
  \frac{K_m(kR)K'_m(kR)}{I_m(kR)K'_m(kR) - \frac{\epsilon^{\infty}_1}{\epsilon_2}
    I'_m(kR)K_m(kR) }.
\end{equation}
Note that the distortion of the Coulomb interaction $V_{\rm ee}(\bm{r},
\bm{r}')$ by the dielectric mismatch between the nanowire and the environment
has been taken into account, which is not addressed in previous work.

$G[ n(\bm{r}) ]$ represents the exchange, correlation and internal energy of the
electrons, which is approximated by the Thomas-Fermi functional
\begin{eqnarray}
  G[ n(\bm{r}) ] & = & \frac{3}{10 m^{\ast}} (3\pi^2)^{2/3} n^{5/3}(\bm{r}),
  \label{app1:eq2}  
\end{eqnarray}
with the exchange-correlation contributions neglected.

From the above Hamiltonian, up to the first order of the perturbation, one can
derive linearized hydrodynamic equations as
\begin{eqnarray}
  \label{app1:eq5}
  \partial_t n_1(\bm{r}) & = & \nabla \cdot [ n_0(\bm{r}) \nabla \eta_1(\bm{r})  ], \\
  \partial_t \eta_1(\bm{r}) & = & \frac{e}{m^{\ast}} \int d\bm{r}' V_{\rm
    ee}(\bm{r}, \bm{r}') n_1(\bm{r}') \nonumber\\
  &&{}+ \frac{5\gamma}{3 m^{\ast}} \frac{n_1(\bm{r})}{[ n_0(\bm{r}) ]^{1/3}},
  \label{app1:eq5_1}
\end{eqnarray}
where $n_1$ and $\eta_1$ are perturbations around the equilibrium. $n_0(\bm{r})
= \bar{n}_0 \Theta(R-\rho)$ is the electron density in the equilibrium with
$\bar{n}_0$ being the average electron density. $\gamma=\frac{(3\pi)^{2/3}}{5
  m^{\ast}}$.

The normal modes from the above equations include both the surface plasmon and
volume plasmon modes. The surface modes, which we focus on in this paper, can be
represented by the ansatz\cite{Perez2009}
\begin{equation}
  \eta_{ q m}(\bm{r}) = R \sum_q \dot{Q}_{ q m} e^{i m \varphi + i q z} \Big[ I_{ m} (q \rho) + A I_{ m} (\kappa \rho) \Big],
  \label{app1:eq3}
\end{equation}
where $I_{m}(\rho)$ is the modified Bessel function of the first kind. $A$
and $\kappa$ are parameters depending on $m$ and $q$, with $m$ being the angular
quantum number of the SPP mode and $q$ representing the SPP wave vector along
the wire axis. Here the first term in the square bracket represents the
incompressible deformation of the electrons while the second term represents the
correction due to the finite compressibility.

By substituting the ansatz Eq.~\eqref{app1:eq3} into Eqs.~\eqref{app1:eq5}
and~\eqref{app1:eq5_1}, one obtain the equation for the parameter $\kappa$
\begin{eqnarray}
  &&1 - \frac{\beta^2}{\omega^2_p}(\kappa^2-q^2) - (qR) I_{ m+1}(qR) K_m(qR) \nonumber\\
  &&{}\times \Big\{ \mathbb{C}'_{ qm} \Big( \frac{\kappa R}{q R} \frac{I_{ m+1}(\kappa R)}{I_{ m+1}(qR)}
  - \frac{I_m(\kappa R)}{I_m(qR)} \Big) \nonumber\\
  &&\hspace{1cm}{}+ \Big[ \frac{\kappa R}{q R} \frac{I_{ m+1}(\kappa R)}{I_{ m+1}(q R)}
  A (1+\mathbb{C}'_{ qm}) \nonumber\\
  &&{}- A \Big(
  \frac{\kappa R}{q R} \frac{I_{ m+1}(\kappa R)}{I_{ m+1}(q R)} + \frac{I_m(\kappa
    R)}{I_{ m+1}(q R)} \frac{K_{ m+1}(q R)}{K_m(q R)} \Big) \Big]
  \Big\} \nonumber\\
  &&\hspace{6cm}{} = 0, \label{app1:eq8}
\end{eqnarray}
with $\omega_p=\sqrt{\frac{e^2 \bar{n}_0}{m^{\ast} \epsilon^{\infty}_1}}$ being
the bulk plasma frequency and $\beta=\sqrt{ \frac{5\gamma}{3 m^{\ast}}
  \bar{n}^{2/3}_0 }$. The parameter $A$ can be expressed as
\begin{equation}
  \label{app1:eq9_1}
  A = - \frac{I_{ m+1}(qR) (qR) (1+\mathbb{C}'_{ qm}) }{ I_{ m+1}(\kappa R) (\kappa R) \Big( 1 +
    \mathbb{C}'_{ qm} \frac{(qR) I_{ m+1}(qR) I_m(\kappa R)}{(\kappa R) I_{ m+1}(\kappa R)
      I_m(q R)} \Big) },
\end{equation}
where
\begin{equation}
  \label{app1:eq10}
  \mathbb{C}'_{ m}(qR) = \frac{ \frac{\epsilon^{\infty}_1}{\epsilon_2} - 1 }{ 1 +
    \frac{\epsilon^{\infty}_1}{\epsilon_2} \frac{I_{ m+1}(qR)}{I_m(qR)} \frac{K_m(qR)}{K_{ m+1}(qR)} }.
\end{equation}
Once $\kappa$ is obtained, the corresponding eigen-frequency $\Omega_{ qm}$
can be calculated from the equation
\begin{eqnarray}
  \label{app1:eq11}
  &&1 + \beta^2 \frac{\kappa^2 - q^2}{\Omega^2_{ qm}} -
  \frac{\omega^2_p}{\Omega^2_{ qm}}
  \Big\{ 1 - (\frac{\epsilon^{\infty}_1}{\epsilon_2}-1) \mathbb{C}_m(qR) \nonumber\\
  &&{}\times \Big[
  (\kappa R) I_{ m+1}(\kappa R) I_m(qR) - (qR) I_m(\kappa R) I_{ m+1}(q R) \Big] \Big\} \nonumber\\
  &&\hspace{6cm} = 0.
\end{eqnarray}
Note that without the dielectric mismatch ($\epsilon^{\infty}_1=\epsilon_2$),
Eqs.~(\ref{app1:eq8}-\ref{app1:eq11}) agree with the results for the surface
mode from Ref.~\onlinecite{Perez2009}.

The Hamiltonian of the SPP can be obtained by substituting Eq.~\eqref{app1:eq3}
into Eq.~\eqref{app1:eq1} and keeping only terms up to the linear order, which
can be written as
\begin{equation}
  H = \frac{2\pi R^3 m^{\ast} \bar{n}_0}{2} \sum_{ q m} \Big[ | \dot{Q}'_{ q
    m} |^2 + \Omega^2_{ q m} |Q'_{ q m}|^2 \Big],
  \label{app1:eq41}
\end{equation}
where the canonical coordinate $Q'_{ q m}$ has the form
\begin{eqnarray}
  Q'_{ q m} & = & \sqrt{\Big[ A I_{ m}
    (\kappa R) \kappa + I_{ m} (q R) q \Big]} \nonumber\\
  && \times \sqrt{\Big[ A I_{ m} (\kappa R) + I_{ m} (q R) \Big] } Q_{ q m}.
  \label{app1:eq55}
\end{eqnarray} 
Following the canonical quantization procedure, the collective coordinate
$Q'_{ q m}$ can be transformed into
\begin{eqnarray}
  Q'_{ q m} & = & \sqrt{\frac{1}{2\pi R^3 m^{\ast} \bar{n}_0 \Omega_{ q m}}}
  (b_{ q m} + b^{\dagger}_{ -q m}),
  \label{app1:eq16}
\end{eqnarray}
where the annihilation(creation) operator $b_{ q m}$($b^{\dagger}_{ q m}$)
satisfies the boson commutation relation $[b_{ q m}, b^\dagger_{ q' m'}] =
\delta_{ q,q'} \delta_{ m m'}$. Substituting Eq.~\eqref{app1:eq16} into
Eq.~\eqref{app1:eq41}, the Hamiltonian of the SPP is quantized as
\begin{eqnarray}
  H_{\rm SPP} & = & \sum_{ q m} \frac{\Omega_{ q m}}{2} (2
  b^{\dagger}_{ q m} b_{ q m} + 1).
  \label{app1:eq17}
\end{eqnarray}
By choosing $m=0$, one obtains the Hamiltonian of the axial symmetric SPP mode
under investigation
\begin{eqnarray}
  \label{app1:eq18}
  H_{\rm SPP} & = & \sum_{ q} \Omega_{ q} b^{\dagger}_{ q} b_{ q},
\end{eqnarray}
without taking into account the zero-point energy which is irrelevant for our
study. The angular quantum number $m$ is also omitted without confusion.

Following the similar procedure, the induced potential of the SPP mode can be
quantized as
\begin{equation}
  \label{app1:eq9}
  V^{\rm in}(\rho, \varphi, z) = \int dq \tilde{V}^{\rm in}_q(\rho, \varphi, z) (b_q + b^{\dagger}_{-q}),
\end{equation}
where 
\begin{eqnarray}
  \label{app1:eq20}
  &&\hspace{-0.5cm} \tilde{V}^{\rm in}_q(\rho, \varphi, z) = - \frac{e^2}{2 \pi \epsilon^{\infty}_1} \sqrt{\frac{ \bar{n}_0}{2 \pi m^{\ast} \omega_p}} \frac{e^{i q z}}{\sqrt{\Omega_q}} \nonumber\\
  &&\hspace{-0.5cm}{}\times \frac{I_0(q \rho) \mathcal{C}''_q + A
    I_0(\kappa \rho)}{\sqrt{[A I_1(\kappa R) \kappa R + I_1(qR) qR] [A
      I_0(\kappa R) + I_0(qR)]}}, \nonumber\\
\end{eqnarray}
with
\begin{eqnarray}
  &&\hspace{-0.5cm} \mathcal{C}''_q = - \Big[ (\kappa R) A I_1(\kappa R) + (qR) I_1(qR) \Big] \nonumber\\
  &&{}\times \Big[ K_0(qR) + \mathbb{C}'_{ 0}(qR) K_0(qR) \Big] \nonumber\\
  &&\hspace{-0.5cm} {}+ A \Big[ (\kappa R) I_1(\kappa R) K_0(qR) + (qR) I_0(\kappa
  R) K_1(qR) \nonumber\\
  &&\hspace{-0.5cm}{}+ \mathbb{C}'_0(qR) K_0(qR) \Big(
  (\kappa R) I_1(\kappa R) - (qR) I_0(\kappa R) \frac{I_1(qR)}{I_0(qR)} \Big) \Big]. \nonumber\\
\end{eqnarray}
The SPP-electron coupling Hamiltonian $H_{\rm SPP\text{-}el}$ in
Eq.~\eqref{cl:eq7} can be obtained by combining the single electron states in
Eq.~\eqref{cl:eq6} and the induced potential of the SPP given above. The
corresponding matrix element reads
\begin{equation}
  \label{app1:eq21}
  g^{ nn'}_{ k-k'} = \int \rho d\rho d\varphi dz \psi^{\ast}_{ n k}(\rho,
  \varphi, z) \tilde{V}^{ in}_{ k-k'}(\rho, \varphi, z) \psi_{ n' k'}(\rho, \varphi, z),
\end{equation}
with $\psi_{ n k}$ given in Eq.~\eqref{cl:eq6}.

Now let us discuss the matrix elements for the interaction Hamiltonians for
electrons. For the ei interaction, we have assumed that the impurities are
distributed on the surface of the nanowire with an axially symmetric
distribution.\cite{Dayeh2007} Given the electron wavefunction $\psi_{ n k}$ in
Eq.~\eqref{cl:eq6}, the corresponding matrix element $v^{ nn'}_q$ can be
calculated as
\begin{eqnarray}
  \label{app1:eq30}
  v^{ nn'}_q & = & \frac{e^2 \delta_{\tilde{m}\tilde{m}'}}{2 \pi^2 \epsilon^{\infty}_1 R} \int^1_0 d\bar{\rho}
  \bar{\rho} \frac{J_{ \tilde{m}}(\lambda^{\tilde{m}}_{\tilde{n}} \bar{\rho})
    J_{ \tilde{m}'}(\lambda^{\tilde{m}'}_{\tilde{n}'} \bar{\rho})}{J_{
      \tilde{m}+1}(\lambda^{\tilde{m}+1}_{\tilde{n}}) J_{
      \tilde{m}'+1}(\lambda^{\tilde{m}'+1}_{\tilde{n}'})} \nonumber\\
  &&\hspace{-1.5cm} \times \Big[ I_0(q R \bar{\rho}) K_0(q R) +
  \mathbb{C}'_0(qR) I_0(q R \bar{\rho}) K_0(qR) \Big].
\end{eqnarray}
Note that the dielectric mismatch effect has been taken into account in the
above expression. For the ep interaction, we here consider the polar interaction
between the electrons and the LO phonons. The corresponding matrix element reads
\begin{eqnarray}
  \label{app1:eq31}
  M^{ nn'}_{ \bm{Q} q} & = & \frac{\sqrt{2 \pi e^2 \Omega_{\rm LO}
      (1/\epsilon^{\infty}_1 - 1/\epsilon^0_1) }}{
    J_{\tilde{m}+1}(\lambda^{\tilde{m}}_{\tilde{n}})
    J_{\tilde{m}'+1}(\lambda^{\tilde{m}'}_{\tilde{n}'})} \frac{2}{|\bm{q}|}
  \nonumber\\
  &&\hspace{-1.5cm} \times \int^1_0 d\bar{\rho} \bar{\rho} J_{\tilde{m}}(\lambda^{\tilde{m}}_{\tilde{n}}
  \bar{\rho}) J_{\tilde{m}'}(\lambda^{\tilde{m}'}_{\tilde{n}'} \bar{\rho}) e^{i
    \bar{\rho} \bm{Q} \cdot \bm{e}_{\rho} R} \delta_{\tilde{m}\tilde{m}'},
\end{eqnarray}
where $\bm{q}=(\bm{Q}, q)$ represents the LO phonon wave vector and $\Omega_{\rm
  LO}$ is the LO phonon energy. $\bm{e}_{\rho}$ is the unit vector along the
radial direction of the wire and $\epsilon^0_1$ is the dielectric constant of
the nanowire in the static limit. Note that here we model the LO phonons by 3D
modes, which is adequate for nanowires with large diameter.\cite{Konar2007,
  Hormann2011} For the ee Coulomb interaction, the corresponding matrix
element reads
\begin{eqnarray}
  \label{app1:eq32}
  V_{q}^{ nn'} & = & \frac{2 e^2}{ \pi \epsilon^{\infty}_1} \int^1_0 d\bar{\rho}_1
  \bar{\rho}_1 \int^1_0 d\bar{\rho}_2 \bar{\rho}_2 \Big[
  I_0(q R \bar{\rho}_{<}) K_0(q R \bar{\rho}_{>}) \nonumber\\
  &&\hspace{-1cm} {}+ ( \frac{\epsilon^{\infty}_1}{\epsilon_2} -1 ) I_0(q R
  \bar{\rho}_1) I_0(q R \bar{\rho}_2) \mathbb{C}_0(qR,\frac{\epsilon^{\infty}_1}{\epsilon_2}) \Big] \nonumber\\
  &&\hspace{-1cm} \times \Big[ \frac{J_{\tilde{m}}(\lambda^{\tilde{m}}_{\tilde{n}}
    \bar{\rho}_1)}{J_{\tilde{m}+1}(\lambda^{\tilde{m}}_{\tilde{n}})} \frac{J_{\tilde{m}'}(\lambda^{\tilde{m}'}_{\tilde{n}'}
    \bar{\rho}_2)}{J_{\tilde{m}'+1}(\lambda^{\tilde{m}'}_{\tilde{n}'})} \Big]^2,
\end{eqnarray}
where $\rho_{>}=\text{max}(\rho_1, \rho_2)$ and $\rho_{<}=\text{min}(\rho_1,
\rho_2)$. Here we have omitted the interband term of the Coulomb interaction
which can be small for nanostructures.\cite{Ell1990} Note that the
  dielectric mismatch effect has been taken into account in the ee Coulomb
  interaction.

\section{Derivation of kinetic equations}
\label{app2}

The time evolution of $B_q(t) = \langle b_q(t) \rangle$ can be derived from the
Heisenberg equation of the annihilation oeprator $b_q$, which reads
\begin{equation}
  \label{app2:eq_01}
  \partial_t B_q(t) = - i \Omega_q B_q(t) + \sum_{ nn', kk', \sigma} p^{
    Q_s}_{ k-k'}  g^{ nn'}_{ k-k'} G^{<}_{ \sigma}(n'k', nk; tt).
\end{equation}
By using the definitions of $\bar{B}_s$ and $P$ given below
Eq.~\eqref{eom:eq4_3}, one has
\begin{eqnarray}
  \label{app2:eq_001}
  \partial_t \bar{B}_s(t) & = & -i \sum_q (\Omega_{ Q_s+q} - \Omega_s) p^{ Q_s}_{ q} B_{ Q_s+q} e^{i \Omega_s t} \nonumber\\
  &&\hspace{-1cm} {}-i \sum_{ nn'\sigma} \sum_{ kk'} p^{ Q_s}_{ k-k'}
  g^{ nn'}_{ k-k'} \Big[ P_{ \sigma}(nk, n'k'; t) \Big]^{\dagger},
\end{eqnarray}
where the first term in the right hand side of the equation describes the effect
of the SPP dispersion on the dynamics of the SPP wave packet and the second term
describes the dissipative effect due to the coupling to electrons. For the SPP
wave packet with sufficiently large length $L$, the line-shape function
$p^{Q_s}_q$ exhibits a sharp symmetric peak around the central wave vector
$Q_s$. In this case, one can perform the Taylor expansion of the dispersive
relation $\Omega_{ Q_s+q}$ around $Q_s$ up to the linear order of $q$ to
eliminate the first term. In doing so, we omit the effect of the SPP dispersion
on the wave packet dynamics and consider only the dissipative effect.

Within the nonequilibrium Green-function approach, the quantum kinetic equation
of electrons can be derived following the Kadanoff-Baym method,\cite{Haug1996,
  mwu2010} yielding
\begin{eqnarray}
  &&\hspace{-1cm} \Big[ -i \partial_t - ( \varepsilon^{n'}_{k'} - \varepsilon^n_k ) \Big]
  G^{\gtrless}_{ \sigma}(nk, n'k'; tt) \nonumber\\
  &&\hspace{0.0cm} = \sum_{ \bar{n} q} ( B_{-q} + B^{\dagger}_q ) \Big[
  g^{ \bar{n} n'}_q G^{\gtrless}_{ \sigma}(nk, \bar{n}k'-q; tt) \nonumber\\
   &&\hspace{0.0cm}\mbox{} - g^{ n \bar{n}}_q G^{\gtrless}_{ \sigma}(\bar{n}k+q, n'k'; tt)
  \Big] + I^{ \gtrless \sigma}_{ nk, n'k'}(t),
  \label{app2:eq1}
\end{eqnarray}
where the first term in the right hand side of the equation is the coherent
driving term of electrons by the SPP and the second term describes the
scattering
\begin{eqnarray}
  I^{ \gtrless \sigma}_{ nk, n'k'}(t) & = & \int^t_{-\infty} d\bar{t} \sum_{\bar{k}}
  \Big\{ \Big[ \Sigma^{\lessgtr}(k \bar{k}; t\bar{t}) G_{\sigma}^{\gtrless}(\bar{k} k';
  \bar{t} t) \nonumber\\
  &&\hspace{-1.5cm}\mbox{}- G^{\gtrless}_{\sigma}(k \bar{k}; t\bar{t})
  \Sigma^{\lessgtr}(\bar{k} k'; \bar{t} t)) \Big] - \Big[ > \longleftrightarrow < \Big] \Big\},
  \label{app2:eq2}
\end{eqnarray}
where $[ > \longleftrightarrow < ]$ stands for the same term as in the previous
$[~]$ but with the interchange of $> \longleftrightarrow <$. It should be
emphasized that due to the driving of the SPP, the dressed Green function
$G^{\gtrless}_{ \sigma}$ has off-diagonal components with respect to the
momentum $k$ and subband index $n$.

By treating the driving term as a perturbation, the kinetic equation can be
linearized by keeping only terms up to the linear order,
\begin{eqnarray}
  &&\hspace{-1cm} \Big[ -i \partial_t - ( \varepsilon^{n'}_{k'} - \varepsilon^n_k ) \Big]
  G^{\gtrless}_{ \sigma}(nk, n'k'; tt) \nonumber\\
  &&\hspace{0.0cm} = ( B_{k-k'} + B^{\dagger}_{k'-k} ) \Big[ g^{ nn'}_{k'-k}
  G^{\gtrless}_{ 0 \sigma}(nk, nk; tt) \nonumber\\
  &&\hspace{0.0cm} {}- g^{ nn'}_{k'-k} G^{\gtrless}_{ 0 \sigma}(n'k', n'k'; tt) \Big] + I^{ \gtrless \sigma}_{ nk, n'k'}(t).
  \label{app2:eq3}
\end{eqnarray}
The first term in the right hand side of the equation is the linearized driving
term. Note that within the linearization, the dressed Green function
$G^{\gtrless}_{ \sigma}$ in the driving term is replaced by the free electron
Green function $G^{\gtrless}_{ 0 \sigma}$, which is diagonal with respect to the
momentum $k$ and subband index $n$. By using the definition $P_{ \sigma}(nk,
n'k'; t) = -i G^{<}_{ \sigma}(nk, n'k'; tt) e^{i \Omega_s t}$ and $f_{
  n\sigma}(k)= -i G^{<}_{ 0\sigma}(nk, nk; tt)$, one obtains
Eq.~\eqref{eom:eq4_2} from the above equation.

Now we turn to the scattering term. We first discuss the ei
scattering. Following the standard approximation,\cite{Haug1996} the
corresponding scattering term can be expressed as
\begin{eqnarray}
  && I^{\rm ei,\sigma}_{ nk, n'k'} = - \Big[ S_{\rm ei, \sigma}(nk, n'k'; >, <) -
  S^{\dagger}_{\rm ei,\sigma}(n'k', nk; <, >) \Big] \nonumber\\
  &&\hspace{1.4cm} \mbox{}+ \Big[ > \longleftrightarrow < \Big], \label{app2:eq7_00}\\
  && S_{\rm{ei},\sigma}(nk, n'k'; >, <) = \int^t_{- \infty} d\bar{t}
  \sum_{ \bar{n} \bar{n}' \tilde{n}} \sum_{ \tilde{k} q} n_i \tilde{v}^{ n \bar{n}}_q
  \tilde{v}^{ \tilde{n} \bar{n}'}_q \nonumber\\
  &&\hspace{0.8cm} {}\times G^{>}_{ \sigma}(\bar{n} k-q, \bar{n}' \tilde{k}-q; t\bar{t})
  G^{<}_{\sigma}(\tilde{n} \tilde{k}, n'k'; \bar{t}t),
  \label{app2:eq7_0}
\end{eqnarray}
where $\tilde{v}^{ nn'}_q=v^{ nn'}_q/\epsilon^{nn'}(q)$ is the screened
electron-impurity interaction. The screening $\epsilon^{nn'}(q)$ is evaluated
following Ref.~\onlinecite{Thoai} in the static limit, which can be written as
\begin{eqnarray}
  \label{app2:eq7_1}
  \epsilon^{nn'}(q) = 1 - \sum_{l\sigma} \frac{V^{nl}_q V^{ln'}_q}{V^{nn'}_q} \sum_k
  \frac{f_{l\sigma}(k+q) - f_{l\sigma}(k)}{\varepsilon^l_{k+q} -
    \varepsilon^l_k - i 0^{+}}.
\end{eqnarray} Note that $S_{\rm ei,\sigma}$ contains the product of two dressed Green
functions. By keeping only terms up to the linear order, $S_{\rm ei,\sigma}$
can be linearized as
\begin{eqnarray}
  &&\hspace{-0.8cm}S_{\rm{ei},\sigma}(nk, n'k'; >, <) = \int^t_{- \infty} d\bar{t}
  \sum_{ \bar{n} \tilde{n}} \sum_{ \tilde{k} q} n_i \tilde{v}^{ n \bar{n} }_q
  \tilde{v}^{ \tilde{n} \bar{n}}_q \nonumber\\
  &&\hspace{0.0cm} {}\times G^{>}_{ 0 \sigma}(\bar{n} k-q, \bar{n} k-q; t\bar{t})
  G^{<}_{\sigma}(\tilde{n} k, n'k'; \bar{t}t) \nonumber\\
  &&\hspace{1.7cm} \mbox{}+ \int^t_{- \infty} d\bar{t}
  \sum_{ \bar{n} \bar{n}'} \sum_{ \tilde{k} q} n_i \tilde{v}^{ n \bar{n} }_q
  \tilde{v}^{ n' \bar{n}'}_q \nonumber\\
  &&\hspace{0.0cm} {}\times G^{>}_{\sigma}(\bar{n} k-q, \bar{n}' k'-q; t\bar{t})
  G^{<}_{ 0 \sigma}(n'k', n'k'; \bar{t}t).
  \label{app2:eq7}
\end{eqnarray}
Note that within the linearization, $S_{\rm ei,\sigma}$ is separated into two
terms. Each term contains only one dressed Green function $G^{\gtrless}_{
  \sigma}$, while the other one is replaced by the free Green function
$G^{\gtrless}_{0 \sigma}$. By applying the rotating wave approximation with
respect to the SPP frequency $\Omega_s$, one can remove the
non-frequency-matched terms in the above equation, yielding
\begin{eqnarray}
  S_{\rm ei, \sigma}(nk, n'k'; >,<) e^{-i \Omega_s t} & = & \pi n_i \sum_{ \bar{n} \bar{n}' q}
  \delta(\varepsilon^{\bar{n}}_{k-q} - \varepsilon^{n'}_{k'} + \Omega_s)
  \nonumber\\
  &&\hspace{-3.3cm}\mbox{} \times \Big[ v^{ n \bar{n}}_q v^{ \bar{n}
    \bar{n}'}_q f^{>}_{ \bar{n} \sigma}(k-q)
  P_{\sigma}(\bar{n}'k, n'k') \nonumber\\
  &&\hspace{-2.9cm}\mbox{}- v^{ n \bar{n}}_q v^{ \bar{n}' n'}_q
  f^{<}_{ n' \sigma}(k') P_{\sigma}(\bar{n} k-q,
  \bar{n}' k'-q) \Big]
  \label{app2:eq8}
\end{eqnarray}
in the Markovian limit. Substituting Eq.~\eqref{app2:eq8} into
Eqs.~\eqref{app2:eq7_00} and~\eqref{app2:eq7_0}, and using the definition
$\bar{I}^{\sigma}_{ nk, n'k'} = I^{\sigma}_{ nk, n'k'} e^{i \Omega_s t}$, one
gets the final expression of the impurity scattering term
\begin{eqnarray}
  &&\hspace{-0.5cm} \bar{I}^{\rm ei,\sigma}_{ nk, n'k'} = \pi n_i \sum_{ \bar{n} \bar{n}' q} \Big\{
  \delta(\varepsilon^{\bar{n}}_{k-q} - \varepsilon^{n'}_{k'}+\Omega_s)
  \nonumber\\
  &&\hspace{-0.5cm}{}\times \Big[ \tilde{v}^{ n
    \bar{n}}_q \tilde{v}^{ \bar{n}'n'}_q P_{\sigma}(\bar{n} k-q, \bar{n}' k'-q) - \tilde{v}^{ n
    \bar{n}}_q \tilde{v}^{ \bar{n} \bar{n}'}_q P_{\sigma}(\bar{n}'k, n'k') \Big] \nonumber\\
  &&\hspace{2.0cm}{}+ \delta(\varepsilon^{\bar{n}'}_{k'-q} - \varepsilon^{n}_{k} -
  \Omega_s) \nonumber\\
  &&\hspace{-0.5cm}{}\times \Big[ \tilde{v}^{ n
    \bar{n}}_q \tilde{v}^{ \bar{n}'n'}_q P_{\sigma}(\bar{n} k-q, \bar{n}' k'-q) - \tilde{v}^{ \bar{n}
    \bar{n}'}_q \tilde{v}^{ \bar{n}' n'}_q P_{\sigma}(nk, \bar{n}k') \Big]
  \Big\}. \nonumber\\
  \label{eom:eq5}
\end{eqnarray}
The ep scattering term can be derived following the similar procedure, which
reads
\begin{eqnarray}
  &&\hspace{0.0cm} \bar{I}^{\rm{ph},\sigma}_{ nk, n'k'} = \pi \sum_{ \bar{n}
    \bar{n}'} \sum_{ q \bm{Q} } \Big\{ \delta( \varepsilon^{\bar{n}}_{k+q} -
  \varepsilon^{n'}_{k'} + \Omega_s + \Omega_{\rm{LO}} ) \nonumber\\
  &&\hspace{0.0cm}{}\times \Big[ M^{n
    \bar{n}}_{\bm{Q} q} M^{\bar{n}' n'}_{\bm{Q} q}
  \Big( N^{>}_{\rm LO} f^{<}_{ n'\sigma}(k') + N^{<}_{\rm LO} f^{>}_{
    n'\sigma}(k')  \Big) \nonumber\\
  &&\hspace{3cm}{}\times P_{\sigma}(\bar{n} k+q, \bar{n}' k'+q) \nonumber\\
  &&\hspace{0.0cm}{}- M^{n \bar{n}}_{\bm{Q} q} M^{\bar{n} n'}_{\bm{Q} q}
  \Big( N^{>}_{\rm LO} f^{>}_{ \bar{n}\sigma}(k+q) + N^{<}_{\rm LO}
  f^{<}_{ \bar{n}\sigma}(k+q) \Big) \nonumber\\
  &&\hspace{3cm}{}\times P_{\sigma}(\bar{n}' k, n'k') \Big] \nonumber\\
  &&\hspace{3.0cm} {}+ \delta( \varepsilon^{\bar{n}}_{{k+q}} - \varepsilon^{n'}_{k'} +
  \Omega_s - \Omega_{\rm{LO}} ) \nonumber\\
  &&\hspace{0.0cm}{}\times \Big[ M^{n \bar{n}}_{\bm{Q} q}
  M^{\bar{n}' n'}_{\bm{Q} q} \Big( N^{<}_{\rm LO} f^{<}_{
    n'\sigma}(k') + N^{>}_{\rm LO} f^{>}_{ n'\sigma}(k') \Big) \nonumber\\
  &&\hspace{3cm}{}\times P_{\sigma}(\bar{n} k+q, \bar{n}' k'+q) \nonumber\\
  &&\hspace{0.0cm}{}- M^{n \bar{n}}_{\bm{Q} q} M^{\bar{n} n'}_{\bm{Q} q} \Big( N^{<}_{\rm LO} f^{>}_{ \bar{n}\sigma}(k+q) +
  N^{>}_{\rm LO} f^{<}_{ \bar{n}\sigma}(k+q) \Big) \nonumber\\
  &&\hspace{3cm}{}\times P_{\sigma}(\bar{n}' k, n'k') \Big] \nonumber\\
  &&\hspace{3.0cm}{}+ \delta( \varepsilon^{\bar{n}'}_{{k'+q}} - \varepsilon^{n}_{k} -
  \Omega_s + \Omega_{\rm{LO}} ) \nonumber\\
  &&\hspace{0.0cm}{}\times \Big[ M^{n \bar{n}}_{\bm{Q} q} M^{\bar{n}' n'}_{\bm{Q} q} \Big( N^{>}_{\rm LO} f^{<}_{ n\sigma} (k) + N^{<}_{\rm
    LO} f^{>}_{ n \sigma} (k) \Big) \nonumber\\
  &&\hspace{3cm}{}\times P_{\sigma}(\bar{n} k+q, \bar{n}' k'+q) \nonumber\\
  &&\hspace{0.0cm}{}- M^{n' \bar{n}'}_{\bm{Q} q} M^{\bar{n}
    \bar{n}'}_{\bm{Q} q} \Big( N^{>}_{\rm LO} f^{>}_{
    \bar{n}'\sigma}(k'+q) + N^{<}_{\rm LO} f^{<}_{ \bar{n}'\sigma}(k'+q)
  \Big) \nonumber\\
  &&\hspace{3cm}{}\times P_{\sigma}(n k, \bar{n} k') \Big] \nonumber\\
  &&\hspace{2.0cm}{}+ \delta( \varepsilon^{\bar{n}'}_{{k'+q}} - \varepsilon^n_k -
  \Omega_s - \Omega_{\rm{LO}} ) \nonumber\\
  &&\hspace{0.0cm}{}\times \Big[ M^{n \bar{n}}_{\bm{Q} q} M^{\bar{n}' n'}_{\bm{Q} q} \Big( N^{<}_{\rm LO} f^{<}_{ n
    \sigma} (k) + N^{>}_{\rm LO} f^{>}_{ n\sigma} (k) \Big) \nonumber\\
  &&\hspace{3cm}{}\times P_{\sigma}(\bar{n} k+q, \bar{n}' k'+q) \nonumber\\
  &&\hspace{0.0cm}{}- M^{n' \bar{n}'}_{\bm{Q} q} M^{\bar{n}
    \bar{n}'}_{\bm{Q} q} \Big( N^{<}_{\rm LO} f^{>}_{
    \bar{n}'\sigma}(k'+q) + N^{>}_{\rm LO} f^{<}_{ \bar{n}'\sigma}(k'+q)
  \Big) \nonumber\\
  &&\hspace{3cm}{}\times P_{\sigma}(n k, \bar{n} k') \Big] \Big\}, 
  \label{eom:eq6}
\end{eqnarray}
where $N^{<(>)}_{\rm LO} = N_{\rm LO} + \frac{1}{2} -(+) \frac{1}{2}$ with
$N_{\rm LO} = 1/[ \exp(\frac{\Omega_{\rm LO}}{k_B T}) - 1 ]$ representing the
thermal LO phonon distribution. The ee scattering term can also be derived in
the similar way,
\begin{eqnarray}
  \bar{I}^{\rm{ee},\sigma}_{ nk, n'k'} & = & 2 \pi \sum_{ q \bar{n} \bar{k}}
  \Big\{ \delta( \varepsilon^n_{k-q} - \varepsilon^{n'}_{k'} + \Omega_s + \varepsilon^{\bar{n}}_{\bar{k} + q} -
  \varepsilon^{\bar{n}}_{\bar{k}} ) \nonumber\\
  &&\hspace{-1.0cm}{}\times \Big[ f^{<}_{ n'\sigma}(k')
  \Pi^{n}_{n'} (\bar{n}\bar{k}, q) \mathit{P_{\sigma}(n k-q, n' k'-q)}
  \nonumber\\
  &&\hspace{0.0cm}\mbox{}- f^{>}_{ n\sigma}(k-q) \Pi^{n}_{n}
  (\bar{n}\bar{k}, q) \mathit{P_{\sigma}(nk, n'k')} \Big] \nonumber\\
  &&\hspace{0.9cm}\mbox{}+ \delta( \varepsilon^n_{k-q} -
  \varepsilon^{n'}_{k'} + \Omega_s + \varepsilon^{\bar{n}}_{\bar{k}} -
  \varepsilon^{\bar{n}}_{\bar{k}-q} ) \nonumber\\
  &&\hspace{-1.0cm}{}\times \Big[ f^{>}_{ n'\sigma}(k')
  \Pi^{n'}_{n} (\bar{n}\bar{k}, -q) \mathit{P_{\sigma}(n k-q, n' k'-q)}
  \nonumber\\
  &&\hspace{0.0cm}\mbox{}- f^{<}_{ n\sigma}(k-q) \Pi^{n}_{n}
  (\bar{n}\bar{k}, -q) \mathit{P_{\sigma}(nk, n'k')} \Big] \nonumber\\
  &&\hspace{0.9cm}\mbox{}+ \delta( \varepsilon^n_{k} -
  \varepsilon^{n'}_{k'-q} + \Omega_s + \varepsilon^{\bar{n}}_{\bar{k}} -
  \varepsilon^{\bar{n}}_{\bar{k}+q} ) \nonumber\\
  &&\hspace{-1.0cm}{}\times \Big[ f^{<}_{ n\sigma}(k)
  \Pi^{n}_{n'} (\bar{n}\bar{k}, q) \mathit{P_{\sigma}(n k-q, n'k'-q)}
  \nonumber\\
  &&\hspace{0.0cm}\mbox{}-  f^{>}_{ n'\sigma}(k'-q) \Pi^{n'}_{n'}
  (\bar{n}\bar{k}, q) \mathit{P_{\sigma}(n k, n'k')}
  \Big] \nonumber\\
  &&\hspace{0.9cm}\mbox{}+ \delta( \varepsilon^n_{k} -
  \varepsilon^{n'}_{k'-q} + \Omega_s + \varepsilon^{\bar{n}}_{\bar{k}-q} -
  \varepsilon^{\bar{n}}_{\bar{k}} ) \nonumber\\
  &&\hspace{-1.0cm}{}\times \Big[ f^{>}_{ n\sigma}(k)\Pi^{n'}_{n}
  (\bar{n}\bar{k}, -q) \mathit{P_{\sigma}(n k-q, n'k'-q)} \nonumber\\    
  &&\hspace{0.0cm}\mbox{}- f^{<}_{ n'\sigma}(k'-q) \Pi^{n'}_{n'}
  (\bar{n}\bar{k}, -q) \mathit{P_{\sigma}(n k, n'k')} \Big] \Big\}, \nonumber\\
  \label{eom:eq7}
\end{eqnarray}
where $\Pi^{n}_{n'} (\bar{n}\bar{k}, q) = \sum_{\sigma} \tilde{V}^{
  n\bar{n}}_{q} \tilde{V}^{ \bar{n} n'}_q f^{>}_{ \bar{n}\sigma}(\bar{k}+q)
f^{<}_{ \bar{n}\sigma}(\bar{k})$ with
$\tilde{V}^{nn'}_q=V^{nn'}_q/\epsilon^{nn'}(q)$ being the screened ee
interaction.

\section{Scattering-induced frequency modification and decay of the
  polarization}
\label{app3}

The polarization $P$ for the $i$-th resonant pair in the presence of the
scattering term $\bar{I}^{\sigma}_{ nk,n'k'}$ given in Eq.~\eqref{eom3:eq0} can
be solved by treating $\Gamma_i$ as perturbation and solve Eq.~\eqref{eom:eq4_2}
order by order. By assuming
\begin{equation}
  \label{app3:eq0}
  P_{ \sigma}(nk,n'k') = \sum^{\infty}_{j=0} P^{ (j)}_{ \sigma}(nk,n'k'; t),
\end{equation}
one obtains
\begin{eqnarray}
  \label{app3:eq1}
  &&\partial_t P^{ (0)}_{ \sigma}(nk, n'k') = i \omega^{ nn'}_{ kk'} P^{ (0)}_{ \sigma}(nk, n'k'; t) \nonumber\\
  &&{}+ i g^{ nn'}_{ k-k'} p^{Q_s}_{ k-k'} \bar{B}^{\dagger}_s \Big[ f_{ n \sigma}(k) - f_{ n' \sigma}(k') \Big],
\end{eqnarray}
for $0$-th order and 
\begin{eqnarray}
  \label{app3:eq2}
  &&\hspace{-1cm}\partial_t P^{ (j+1)}_{ \sigma}(nk, n'k') = i \omega^{ nn'}_{ kk'} P^{ (j+1)}_{ \sigma}(nk, n'k'; t) \nonumber\\
  &&\hspace{-1cm}{}+ \sum_q \Gamma_{i} [ P^{ (j)}_{\sigma}(nk-q, n'k'-q) - P^{ (j)}_{\sigma}(nk, n'k') ],
\end{eqnarray}
for $(j+1)$-th order, with $j \ge 0$.

In the long time limit $t \to \infty$, the solution of the above equations reads
\begin{eqnarray}
  \label{app3:eq3}
  P^{ (0)}_{\sigma}(nk, n'k') & = & - g^{ nn'}_{ k-k'} p^{Q_s}_{ k-k'}
  \bar{B}^{\dagger}_s \nonumber\\
  &&\hspace{-1.5cm}{} \times [ f_{n\sigma}(k) - f_{n'\sigma}(k') ]/(\omega^{ nn'}_{ kk'} + i 0^{+}),
\end{eqnarray}
for $0$-th order, and 
\begin{eqnarray}
  \label{app3:eq4}
  P^{ (j+1)}_{\sigma}(nk, n'k') & = & \frac{P^{ (j)}_{\sigma}(nk,
    n'k')}{\omega^{ nn'}_{ kk'} + i 0^{+}} \nonumber\\
  &&\hspace{-2cm}{} \times i \Gamma_i \sum_q [ \frac{P^{ (j)}_{\sigma}(nk-q, n'k'-q)}{P^{ (j)}_{\sigma}(nk, n'k')}- 1 ],
\end{eqnarray}
for $(j+1)$-th order.

Now let's evaluate $\sum_q [ \frac{P^{ (j)}_{\sigma}(nk-q, n'k'-q)}{P^{(j)}_{\sigma}(nk, n'k')}- 1 ]$. For $0$-th order,
\begin{eqnarray}
  && \sum_q \Big[ \frac{P^{ (0)}_{\sigma}(nk-q, n'k'-q)}{P^{
      (0)}_{\sigma}(nk, n'k')}- 1 \Big] \nonumber\\
  && = \sum_q \Big[ \frac{ g^{ nn'}_{ k-k'} p^{Q_s}_{ k-k'}
    \bar{B}^{\dagger} [ f_{n\sigma}(k) - f_{n'\sigma}(k') ]}{g^{ nn'}_{
      k-k'} p^{Q_s}_{ k-k'} \bar{B}^{\dagger} [ f_{n\sigma}(k-q) -
    f_{n'\sigma}(k'-q) ] } \nonumber\\
  &&{}\times \frac{\omega^{ nn'}_{ kk'}}{\omega^{ nn'}_{ k-q,k-q'} + i 0^{+}}- 1 \Big] \nonumber\\
  && \approx \sum_q [ \frac{\omega^{ nn'}_{ kk'}}{\omega^{
      nn'}_{ k-q,k-q'} + i 0^{+}}- 1 ],
  \label{app3:eq5}
\end{eqnarray}
where we have assumed that $f_{n\sigma}(k)$ varies slowly with respect to $k$
and can be cancelled out. The $1$-st order term can be evaluated as
\begin{eqnarray}
  &&\sum_q \Big[ \frac{P^{ (1)}_{\sigma}(nk-q, n'k'-q)}{P^{
      (1)}_{\sigma}(nk, n'k')} - 1 \Big] \nonumber\\
  &&{}= \sum_q \Big\{ \sum_{q'} [ P^{
    (0)}_{\sigma}(nk-q-q', n'k'-q-q') \nonumber\\
  &&{}- P^{ (0)}_{\sigma}(nk-q, n'k'-q) ] \{ \sum_q [ P^{ (0)}_{\sigma}(nk-q, n'k'-q) \nonumber\\
  &&{}- P^{ (0)}_{\sigma}(nk, n'k') ] \}^{-1} \frac{\omega^{ nn'}_{ kk'}}{\omega^{ nn'}_{ k-q,k-q'} + i 0^{+}} - 1 \Big\} \nonumber\\
  &&{} \approx \sum_q [ \frac{\omega^{ nn'}_{ kk'}}{\omega^{nn'}_{ k-q,k-q'} + i 0^{+}} - 1 ],  \label{app3:eq6}
\end{eqnarray}
where we have assumed that $\sum_q[ P^{ (0)}_{\sigma}(nk-q, n'k'-q) -
P^{(0)}_{\sigma}(nk, n'k') ]$ varies slowly with respect to $k$($k'$) and can be
cancelled out.

Following the similar procedure, for $j$-th order, one has
\begin{eqnarray}
  && \sum_q [ \frac{P^{ (j)}_{\sigma}(nk-q, n'k'-q)}{P^{
      (j)}_{\sigma}(nk, n'k')}- 1 ] \nonumber\\
  && \approx \sum_q [ \frac{\omega^{ nn'}_{ kk'}}{\omega^{nn'}_{ k-q,k-q'} + i 0^{+}} - 1 ],
  \label{app3:eq7}
\end{eqnarray}
with $j \ge 1$.

By combining
Eqs.~\eqref{app3:eq0},~\eqref{app3:eq3},~\eqref{app3:eq4},~\eqref{app3:eq5}
and~\eqref{app3:eq7}, one obtains Eqs.~(\ref{eom3:eq1}-\ref{eom3:eq1_2}).

\section{Effect of the SO phonons}
\label{app5}

The SO phonons in nanowires\cite{Hormann2011} can also influence the damping of
the SPP. However, for the typical InAs nanowires we considered here, the SO
phonons are of marginal importance since their contribution is much smaller than
that from the LO phonons. This will be shown in this Appendix.

The induced potential due to the SO phonon is given by\cite{Klimin1994,
  Bennett1995, Vartanian2005}
\begin{eqnarray}
  V_{\rm SO} & = & \sum_p ( \Gamma^{\rm SO}_p e^{i q z} e^{i p \varphi} a^{\rm SO}_p + \mathit{h.c.} ),
\end{eqnarray}
with
\begin{eqnarray}
  \Gamma^{\rm SO}_p & = & \sqrt{\frac{2 \pi e^2 R}{q} \frac{D(\Omega_{\rm
        SO})}{I_p(qR)I'_p(qR)}} I_p(qr), \\
  D(\omega) & = & \frac{\epsilon_2(\omega)}{\epsilon_2(\omega)
    \frac{\partial}{\partial \omega} \epsilon_1(\omega) - \epsilon_1(\omega)
    \frac{\partial}{\partial \omega} \epsilon_2(\omega)}, \\
  \epsilon_{i}(\omega) & = & \epsilon^{\infty}_i\frac{\Omega_{\rm Li}^2 -
    \omega^2}{\Omega_{\rm Ti}^2 - \omega^2},
\end{eqnarray}
where $a^{\rm SO}_p$ is the annihilation operator for the SO phonons and
$\Omega_{\rm SO}$ is the SO phonon energy. $\Omega_{\rm Li}$($\Omega_{\rm Ti}$)
represents the LO(TO) phonon energy inside ($i=1$) and/or outside ($i=2$) the
nanowire. The coordinates are chosen according to Fig.~\ref{cl:fig1}(a).

\begin{figure}
  \centering
  \includegraphics[width=5.5cm]{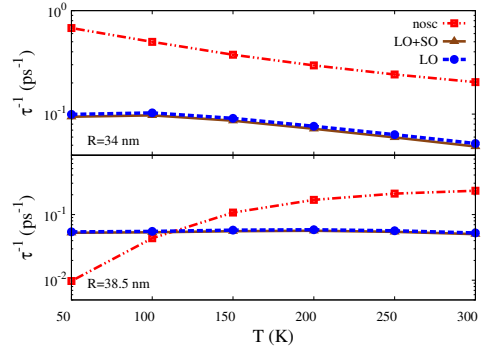}\\
  \caption{Temperature dependence of SPP damping rate in the presence of the ep
    scattering for $R=34$ and $38.5$~nm. Blue dashed curves represent the
    results with only the LO phonon scattering, while the brown solid curves
    represent the results with both the LO and SO phonon scatterings. The
    damping rates without any scattering are plotted with red double-dotted
    chain curves for comparison. Other parameters are all the same as
    Fig.~\ref{re:fig1}.}
  \label{app5:fig1}
\end{figure}

As the SO phonon energy is very close to the LO phonon energy for typical InAs
nanowires,\cite{Hormann2011} we assume $\Omega_{\rm SO} = \Omega_{\rm LO}$ in
the calculation. By using the single electron states in Eq.~\eqref{cl:eq6}, the
corresponding electron-SO-phonon interaction Hamiltonian can be written as
\begin{eqnarray}
  \hspace{-0.8cm}H_{\rm SO} & = & \sum_{p q} \sum_{n n' k \sigma} \bar{M}^{nn'}_{p q} [ a^{\rm SO}_{p q} +
  (a^{\rm SO}_{-p -q})^{\dagger} ] c^{\dagger}_{n k \sigma} c_{n' k-q \sigma},
\end{eqnarray}
where the interaction matrix element reads
\begin{eqnarray}
  \bar{M}^{ nn'}_{p q} & = & \frac{\sqrt{\pi e^2 \Omega_{\rm SO}
      (1/\epsilon^{\infty}_1 - 1/\epsilon^0_1)R/q }}{
    J_{\tilde{m}+1}(\lambda^{\tilde{m}}_{\tilde{n}})
    J_{\tilde{m}'+1}(\lambda^{\tilde{m}'}_{\tilde{n}'})} \delta_{\tilde{m},\tilde{m}'+p}
  \nonumber\\
  &&\hspace{-1.5cm} \times \int^1_0 d\bar{\rho} \bar{\rho} J_{\tilde{m}}(\lambda^{\tilde{m}}_{\tilde{n}}
  \bar{\rho}) J_{\tilde{m}'}(\lambda^{\tilde{m}'}_{\tilde{n}'} \bar{\rho})
  \frac{I_p(qR\bar{\rho})}{\sqrt{I_p(qR) I'_p(qR)}}.
\end{eqnarray}

By using the analytic solution given in Secs.~\ref{sec2_3} and~\ref{sec2_4}, we
calculate the temperature dependence of the SPP damping rates for $R=34$ and
$38.5$~nm in the presence of the ep scattering with and without the contribution
of the SO phonons, which is plotted in Fig.~\ref{app5:fig1}. One can see that
the damping rates are dominated by the LO phonons and the SO phonons have very
little effect.

\section{Visualization of the broadening and shifting}
\label{app4}

We have seen that the SPP damping rate can be understood as the broadening and
shifting of the resonant pairs. In the analytic solution, the broadening and
shifting can be seen clearly from both the resonant denominator in the
polarization [Eq.~\eqref{eom3:eq1}] and the Lorentzian in the damping rate
[Eq.~\eqref{eom3:eq2_2}]. One may wonder whether the broadening and shifting can
also be observed in a more intuitive way from the numerical results. In fact,
inspired by Eq.~\eqref{eom3:eq1}, the broadening and shifting of the resonant
pairs can be visualized from the normalized polarization $P(nk, n'k')/(\delta f
\cdot \bar{B}_s)$, from which the structure of the resonant denominator can be
identified.

\begin{figure}
  \centering
  \includegraphics[width=7.5cm]{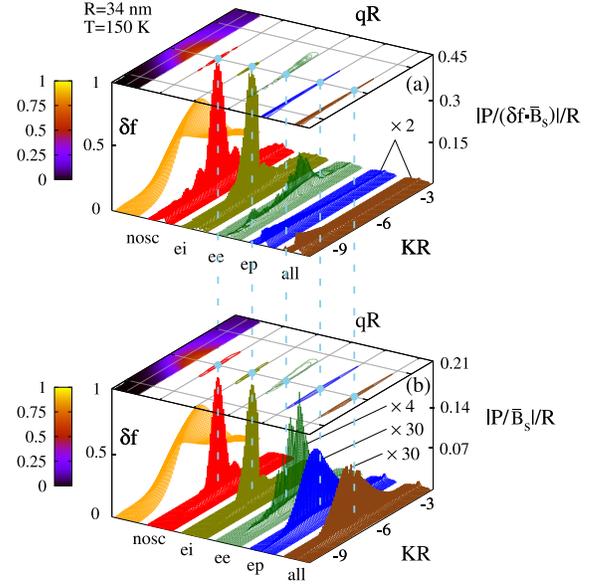}
  \caption{(Color online) 3D plot and the corresponding contour plot of the
    normalized (a) and unnormalized (b) electron polarization corresponding to
    the resonant pair (i) for $R=34$~nm, $T=150$~K. The corresponding population
    difference is also plotted in the figure. The olive, green and blue curves
    represent the polarization with the ei, ee and ep scattering only,
    respectively. The red curves represent the polarization without scattering,
    while the brown curves represent the polarization in the presence of all the
    three scatterings. In the contour plots, the resonance corresponding to the
    SPP central wavevector $Q_s$ is shown with skyblue dot in
    the contour plot. The population difference is plotted with orange
    curves. Different curves are offset along $q$-axis for clarify. To provide a
    clear visualization, in (a) the normalized polarizations with the ep and all
    the scatterings are enlarged by a factor of $2$. In (b), the unnormalized
    polarization with the ee/ep/all scattering is enlarged by a factor of
    $4$/$30$/$30$.}
  \label{re:fig4}
\end{figure}

Before we show the numerical results of the normalized polarizations, we first
explain what we expect from the polarizations. According to the analytic
solution Eq.~\eqref{eom2:eq1}, the magnitude of the normalized polarization
$|P(nk, n'k')/(\delta f \cdot \bar{B}_s)|$ without scattering can exhibit a
sharp Lorentzian peak around the resonance corresponding to the SPP central
wavevector $Q_s$ as indicated by the resonant denominator. Several side peaks
can also exist as the SPP wave packet is nonmonochromatic [e.g., the line-shape
function $p^{Q_s}_q \ne \delta(q-Q_s)$]. With scattering, these peaks can be
shifted and broadened, indicating the shifting and broadening of the
corresponding resonant pairs. Small side peaks may also be smeared out by the
broadening. Note that with the scattering, the polarizations from the numerical
results can have more complex behaviors. The peaks in the polarization can also
be distorted by the scattering, which has been omitted in the analytic solution.

\begin{figure}
  \centering
  \includegraphics[width=7.5cm]{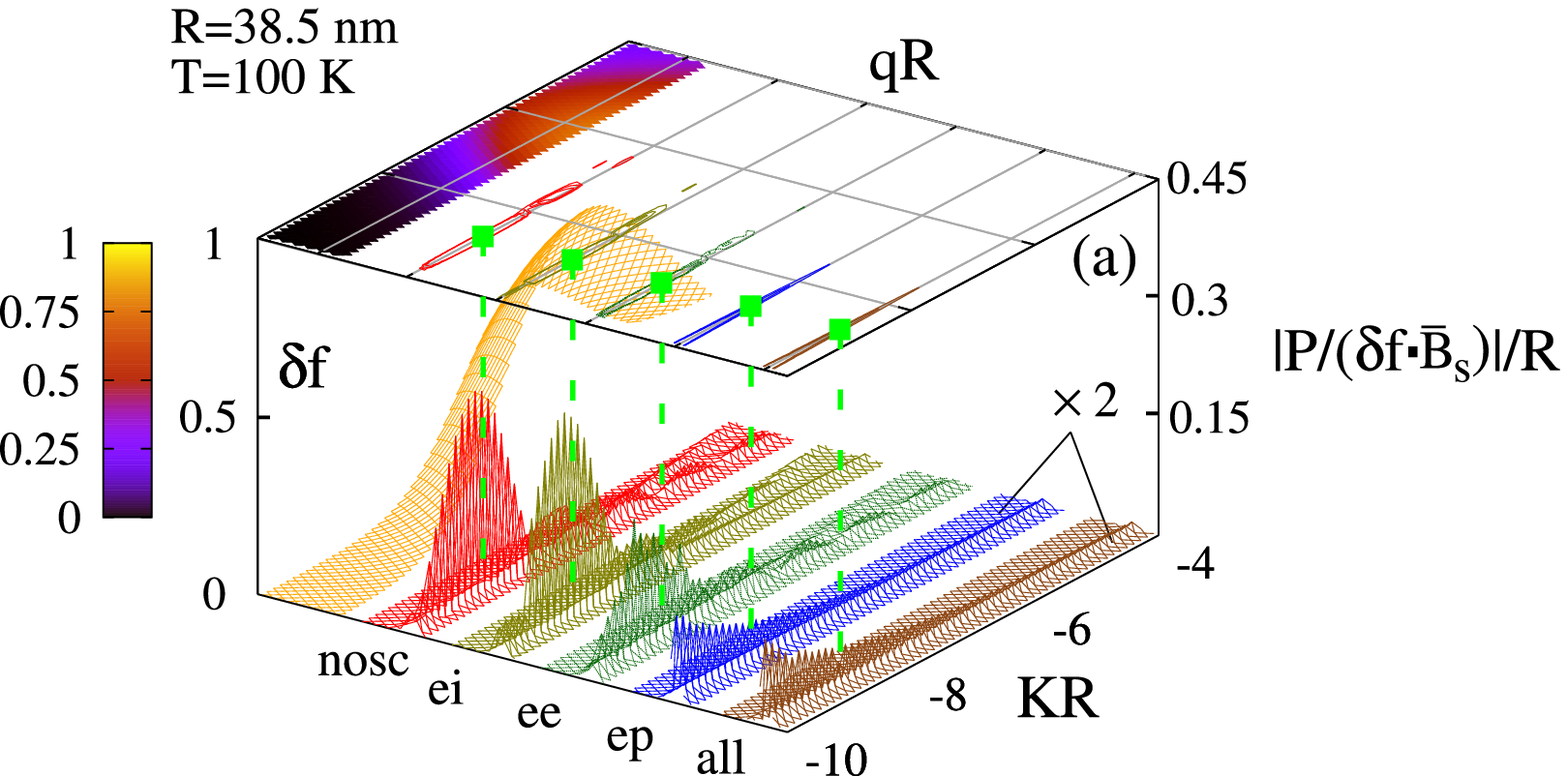}
  \includegraphics[width=7.5cm]{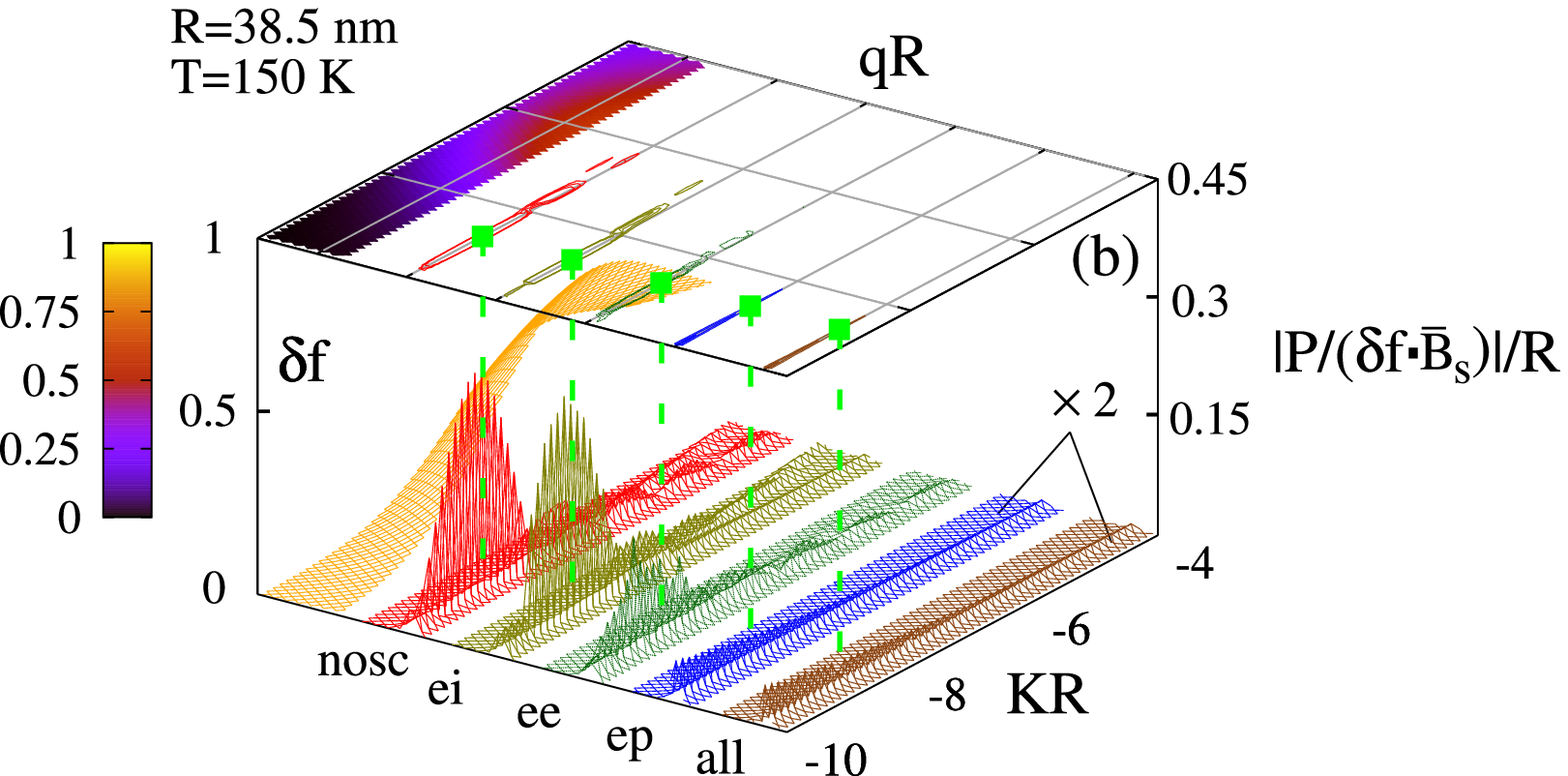}
  \includegraphics[width=7.5cm]{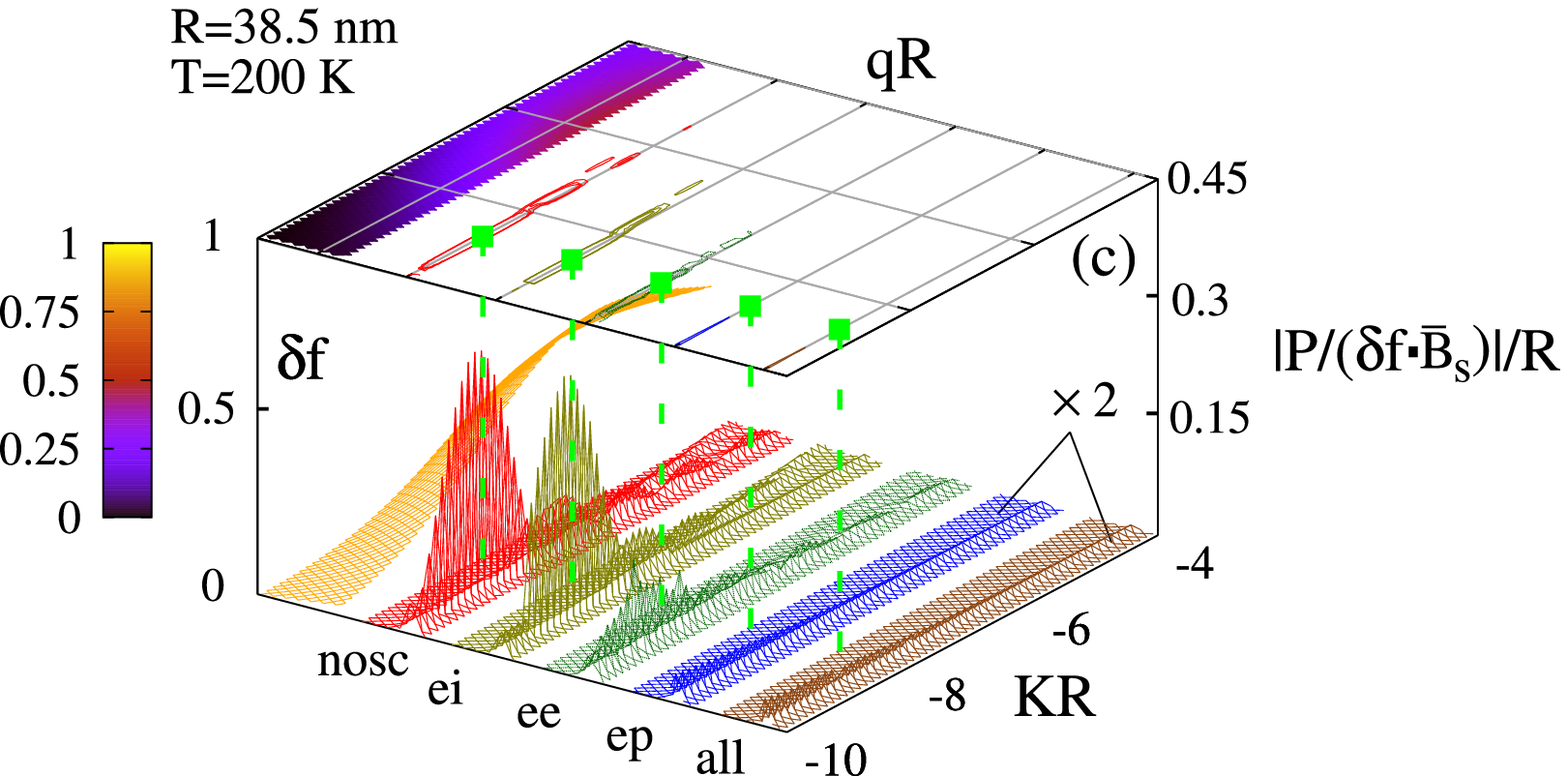}
  \caption{(Color online) 3D plot and the corresponding contour plot of the
    normalized electron polarization corresponding to the resonant pair (ii) for
    (a) $R=38.5$~nm, $T=100$~K, (b) $R=38.5$~nm, $T=150$~K and (c) $R=38.5$~nm,
    $T=200$~K. The corresponding population difference is also plotted in the
    figure. The resonance corresponding to the SPP central wavevector $Q_s$ is
    shown with green square in the contour plot.}
  \label{re:fig5}
\end{figure}

We first concentrate on the typical case with $R=34$~nm and $T=150$~K
corresponding to the strong Landau damping regime. The magnitudes of the
normalized electron polarization $|P(nk, n'k')/(\delta f \cdot \bar{B}_s)|$ from
the computation are plotted in Fig.~\ref{re:fig4}(a). In the figure,
polarizations with different scatterings are plotted by curves with different
colors. The contour plots of the polarizations are also shown in the $K$-$q$
plane, which are useful for identifying the shape and position of the
polarizations. The population differences $\delta f=f_{ n'}(k') - f_{ n}(k)$ for
the resonant pairs are also plotted with orange curves for comparison, with the
corresponding contour map plotted in the $K$-$q$ plane. The polarizations with
different scatterings and the population difference have been offset along the
$q$-axis for clarification. The corresponding unnormalized electron
polarizations $|P(nk, n'k')/\bar{B}_s|$ are also
plotted in the similar way in Fig.~\ref{re:fig4}(b) for comparison. Note that
all the polarizations are taken at time $t=2.86$~ps. Polarizations taken at
other time show similar behaviors.

In Fig.~\ref{re:fig4}(a), the polarization is plotted for the resonant pair (i)
which is the only relevant resonant pair for the damping. The resonant pair is
centralized around the resonance corresponding to the SPP central wavevector
$Q_s$, which is represented by the skyblue dots in Fig.~\ref{re:fig4}(a). One
finds that without scattering, the corresponding polarization (red curve)
exhibits a sharp main peak around the resonance. Several side peaks exist around
the main peak. These features agree with the previous discussion. Note that the
main peak lies in the regime with large $\delta f$, indicating that the SPP
absorption by the polarization is strong. Also note that due to the strong
resonance, the profile of the unnormalized polarization is mainly decided by the
resonance and shares a similar structure as the normalized one.

In the presence of the ei scattering, the polarization is slightly
modified. Some side peaks are smeared out. However, the broadening is rather
small and the main peak is almost unchanged. This can be better seen by
comparing the polarization with the ei scattering (olive curve) to the one
without scattering (red curve). According to the previous discussion, these
features indicate that the ei scattering introduces a small broadening.

In the presence of the ee scattering, a shifting of the peaks can be seen
clearly by comparing the contour plot of the corresponding polarization (green
curve) to the one without scattering (red curve). Note that small side peaks can
also be identified with the ee scattering. These features indicate that the ee
scattering introduces a large shifting to the corresponding resonant pairs. The
broadening due to the ee scattering is small since the small side peaks are not
smeared out. It is also noted that the ee scattering can distort the
polarization. The peak becomes fragmented and the profile becomes
non-Lorentzian. However, as the influence to the SPP damping comes from the
summation of the polarizations as indicated by Eq.~\eqref{eom:eq4_1}, the effect
of these distortion on the SPP damping is marginal.

In the presence of the ep scattering, the peak in the polarization (blue curve)
is markedly broadened and all the side peaks are smeared out, indicating that
the effect of the ep scattering introduces a large broadening. Note that the
maximum of the broadened peak is also shifted compared to the one without
scattering. This shows that the ep scattering can also have contribution to the
shifting of the resonant pairs. However, due to the large broadening, the effect
of such shifting on the SPP damping is marginal. We also point out that due to
the large broadening, the profile of the unnormalized polarization is mainly
determined by the population difference rather than the resonance, which can be
seen from the corresponding unnormalized polarization (blue curve) in
Fig.~\ref{re:fig4}(b).

In the presence of all the three scatterings, one finds that the polarization
(brown curve) has almost the same profile as the one with the ep scattering
only. The broadening and shifting due to the ei and ee scatterings are
negligible due to the large broadening introduced by the ep scattering. This
indicates that the ep scattering plays the dominant role.

For the weak Landau damping regime, as the scattering has different influence on
the SPP damping rate in different temperature regimes. We plot the normalized
polarizations in Fig.~\ref{re:fig5} for three typical cases: (a) $R=38.5$~nm,
$T=100$~K, (b) $R=38.5$~nm, $T=150$~K and (c) $R=38.5$~nm, $T=200$~K,
corresponding to the low, intermediate, and high temperature regimes,
respectively. The polarizations are plotted for the resonant pair (ii) which
dominates the SPP damping in these cases. Similar broadening and shifting can
also be identified in the figure. Note that in these cases, the shifting due to
the ee scattering is rather small and difficult to be identified in the figures.


\begin{thebibliography}{0}
\bibitem{comment2} For sufficiently long SPP wave packet ($L\gtrsim 100
    R$), the results are not sensitive to the wave packet length $L$.
\bibitem{comment3} The influence of the ei scattering depends on both the
  impurity configuration and the impurity density. In addition to the surface
  impurities, we have also performed the calculation for bulk-like impurities
  (i.e., impurities evenly distributed inside the nanowire. The corresponding
  line density of the bulk impurities is set equal to the line density of the
  surface impurities $n_i$). For both impurity configurations, we have checked
  that the influence of the ei scattering remains very small for $n_i< 5 n_e$.
\end{thebibliography}
\end{document}